\documentclass[12pt,notitlepage,a4paper]{article}
\pdfoutput=1
\usepackage{amssymb}
\usepackage{delarray,amsmath,bbm,epsfig}
\usepackage{cite}

\usepackage{color, graphicx} 

\newcommand\fverb{\setbox\pippobox=\hbox\bgroup\verb}
\newcommand\fverbdo{\egroup\medskip\noindent%
                              \fbox{\unhbox\pippobox}\ }
\newcommand\fverbit{\egroup\item[\fbox{\unhbox\pippobox}]}
\newbox\pippobox
\newcommand{\nn}{\nonumber}
\newcommand{\beq} {\begin{equation}}
\newcommand{\eeq} {\end{equation}}
\newcommand{\beqa} {\begin{eqnarray}}
\newcommand{\eeqa} {\end{eqnarray}}

\newcommand{\jsn}{\mathrm{sn}}

\newcommand{\jsc}{\mathrm{sc}}

\newcommand{\jnd}{\mathrm{nd}}
\newcommand{\jns}{\mathrm{ns}}

\newcommand{\ie}{{\it i.e.}}

\newcommand{\eps}{\epsilon}
\newcommand{\ieps}{i\varepsilon}
\newcommand{\order}[1]{${\cal O}\left(#1 \right)$}
\newcommand{\morder}[1]{{\cal O}\left(#1 \right)}
\newcommand{\eq}[1]{(\ref{eq:#1})}

\newcommand{\inv}[1]{\frac{1}{#1}}

\newcommand{\re}{{\rm Re}}

\newcommand{\mC}{\mathcal{C}}

\newcommand{\mE}{\mathcal{E}}

\newcommand{\be}{\begin{equation}}
\newcommand{\ee}{\end{equation}}
\newcommand{\bea}{\begin{eqnarray}}
\newcommand{\eea}{\end{eqnarray}}

\begin{document}
\begin{flushright}
{\it CP$^\textrm{\it 3}$-Origins-2010-9 } \\
HIP-2010-05/TH
\end{flushright}
\vskip 2cm \centerline{\Large {\bf Electrostatics of Coulomb gas, 
}} \vskip 3mm
\centerline{\Large {\bf lattice paths, and discrete polynuclear growth}} \vskip 1cm
\renewcommand{\thefootnote}{\fnsymbol{footnote}}
\centerline{{\bf Niko
Jokela,$^{1,2}$\footnote{najokela@physics.technion.ac.il} Matti
J\"arvinen,$^{3}$\footnote{mjarvine@cp3-origins.net} and Esko
Keski-Vakkuri$^{4,5}$\footnote{esko.keski-vakkuri@helsinki.fi}}}
\vskip .5cm \centerline{\it ${}^{1}$ Department of Physics}
\centerline{\it Technion, Haifa 32000, Israel}
\centerline{\it
${}^{2}$ Department of Mathematics and Physics}
\centerline{\it University of Haifa at Oranim}
\centerline{\it Tivon 36006, Israel}
\centerline{\it ${}^{3}$CP$^\textrm{\it 3}$-Origins, 
}
\centerline{\it Campusvej 55, DK-5230 Odense M, Denmark}
\centerline{\it
${}^{4}$Helsinki Institute of Physics and ${}^{5}$Department of
Physics } \centerline{\it P.O.Box 64, FIN-00014 University of
Helsinki, Finland}

\setcounter{footnote}{0}
\renewcommand{\thefootnote}{\arabic{footnote}}

\begin{abstract}
We study the partition function of a two-dimensional Coulomb gas on a circle, in the presence of external pointlike charges, in a double scaling limit where both the external charges and the number of gas particles are large. Our original motivation comes from studying amplitudes for multi-string emission from a decaying D-brane in the high-energy limit. We analyze the scaling limit of the partition function and calculate explicit results.
We also consider applications to random matrix theory. The partition functions can be related to
random scattering, or to weights of lattice paths in certain growth models.
In particular, we consider the discrete polynuclear growth model and use our results to compute the cumulative probability density for the height of long level-1 paths. We also obtain an estimate for an almost
certain maximum height.

\end{abstract}

\newpage

\tableofcontents


\section{Introduction}

Unstable D-branes have many decay channels. The simplest, and a relatively well understood one, is closed string emission, one at a time at tree
level \cite{Lambert:2003zr,Gaiotto,Shelton,Senreview}.
The multi- closed string or multi- open string production channels lead to more complicated calculations, and are less well understood.
In a sequence of papers \cite{Jokela:2007yc,npt} we have developed
estimates for amplitudes in the multi- open string channels, and most recently in the closed string pair production channel \cite{Jokela:2009fd}.
Our calculational strategy was based on mapping the problem to calculating the partition function of a log-gas of
unit charges in a unit circle (the Dyson gas \cite{Dyson:1962es}, arising from the background of the condensing tachyons on the brane) with additional test charges (arising
from the vertex operators for string emission) \cite{VEPA,VEPA2,rest}. We then used electrostatics to find the leading saddle-point contribution. This is sufficient for
the high-energy limit of the emission amplitude \cite{Jokela:2009fd}.

Logarithmic Coulomb interaction potential arises as a solution of the Poisson equation in two dimensions. After the original paper \cite{Dyson:1962es}, thermodynamics of more general Coulomb
systems have been studied, {\em e.g.}, in \cite{coulombthermo}. Some applications of Coulomb gas thermodynamics are discussed in \cite{Saleurreview}. Furthermore, Dyson gas with
additional charges has applications in various contexts in random matrix theory (RMT), such as estimates for scattering amplitudes in a quantum mechanical random scattering problem, and estimates for the weights of long lattice paths in growth models.
It is therefore worthwhile to study if the electrostatics approach can
be used to derive explicit results.

In this paper we analyze the electrostatics approach in more detail. We will also discuss
applications to random scattering and to growth models. As a first example we consider the boundary one-point partition function, with one test
charge on the boundary. 
This configuration is solved exactly in the literature, and serves as a test for the electrostatic method.
We point out that the one-point function 
is related to lattice paths in the ``lock step model of vicious walkers'' \cite{Fisherwalk,Forrestwalk,Forrester}. 
We also derive new results for higher point boundary functions. Their application to open string emission from decaying D-branes will be discussed elsewhere.
We then consider bulk $n$-point functions with $n$ external charges inside the unit circle. In string theory, they arise
from (multi) closed string emission from decaying D-branes. A more widely known physics application is multi-channel scattering across a
non-ideal lead (with a tunneling barrier) coupled to a chaotic cavity. We show that the bulk one-point partition function, which was analyzed in detail in our earlier work \cite{Jokela:2009fd}, can be related to the Poisson
kernel associated with the eigenvalue probability distribution function for the scattering matrix.
Another RMT interpretation, applicable for $n$-point functions for all $n\geq1$,  is given by weights of a class of long lattice paths in the
discrete polynuclear growth (PNG) model (we follow the presentation in \cite{png,Forrester}).  
Using the results from our electrostatic analysis in \cite{Jokela:2009fd}, we
give an estimate for the cumulative probability distribution function of long paths from large weighted matrices.
This limit also corresponds to the high-energy limit of closed string pair production from a decaying D-brane \cite{Jokela:2009fd}.

As the above examples hopefully show, there is an interesting connection between D-brane decay in string theory and problems in random
matrix theory. We anticipate that there are many other examples to be found. The connections help in finding useful strategies
for problem-solving in string theory.
They should also serve as additional motivation for work in random matrix theory. For example, in the string theory context one is naturally
lead to consider log-gases in grand canonical ensembles, whereas our impression is that most of mathematical work is focused on canonical
ensembles. A very recent example of potentially related work on grand canonical ensembles is \cite{zabrodin}.

The sectioning of the paper is as follows. In Section \ref{sec:largeN} we introduce our
approach and show that the
leading behavior of the partition functions can be calculated by using continuum electrostatics. 
We also analyze the associated electrostatic potential problem.
In Section~\ref{sec:1pt}, as a first example we calculate in detail the leading behavior of the partition function with one external charge on the unit circle, and compare with the available exact result. We also point out
the lattice path interpretation. In Section~\ref{sec:npts} we move to consider boundary $n$-point functions, analyzing the two-point function in detail. In Section~\ref{sec:b1pt} we move to consider external charges
in the bulk, starting with the one-point function. Depending on the chosen parameters, this case is either
very simple, with application to random scattering, or more complicated but where we can use the electrostatic
approximation in the double scaling limit. Finally in Section~\ref{sec:bulknpt} we consider bulk $n$-point functions
and their relation to the discrete polynuclear growth model, and calculate the cumulative probability density
function in a special case. We also discuss a connection to lattice paths. Appendix \ref{app:ooN} contains calculational details. In \cite{Jokela:2010zp} we apply the results of Section~\ref{sec:npts} to D-brane decay.


\section{Large $N$ expansion of the partition function} \label{sec:largeN}

We consider the partition function\footnote{The full partition function factorizes to $Z=Z_{kin}Z_{pot}$ where the contribution from kinetic energies $Z_{kin}$ involves only simple Gaussian integrals. Hence we focus on
 the non-trivial potential energy contribution only.} of a two-dimensional Coulomb gas on a circle (Dyson gas) with $n$ external charges $\xi_a$ at fixed positions $w_a$ on the complex plane \cite{Jokela:2007yc,npt}:
\be \label{eq:Zredef}
 Z_{n}(\beta;\{\xi_a\};N)  = \frac{1}{N!} \int\prod_{i=1}^N\frac{dt_i}{2\pi} e^{-\beta H} \ ,
\ee
where the Hamiltonian $H$ reads
\be \label{eq:Ham}
 H = - \sum_{1\le i<j\le N} \log |e^{it_i}-e^{it_j}| - \sum_{i=1}^N\sum_{a=1}^n \xi_a \log |e^{it_i}-w_a| - \sum_{1\le a<b \le n} \xi_a \xi_b \log|w_a-w_b| \ \ .
\ee
It describes an ensemble of $N$ unit charges on the unit circle at $e^{it_1},\ldots,e^{it_N}$, which interact via the Coulomb potential and are subject to the electric field created by the fixed test charges $\xi_a$. The charges are located at $w_a$ anywhere on the complex plane. The charges with $|w_a|=1$ ($|w_a|<1$) will be called boundary (bulk) charges. For the boundary charges we also denote $w_a=e^{i\tau_a}$.

A useful strategy is to study the system using classical electrostatics, first in the limit of a large number $N$ of unit charges. In this limit the unit charges can be
treated as a continuous charge distribution. Its density is modified by the external charges $\xi_a$, and we are interested in finding the electrostatic
equilibrium configuration.
In particular, nontrivial electrostatic configurations are obtained when the external charges are of the same order as the sum of the unit charges, $N \sim \xi_a$. We shall in fact study the behavior of the system in a double scaling limit limit $N,\xi_a \to \infty $ with all ratios $N/\xi_a$ fixed.\footnote{The behavior of  $Z_{n}$ in the limit $N \to \infty$ while holding $\xi_a$ fixed
can be calculated using the asymptotics of Toeplitz determinants \cite{VEPA,Fisher}.}

The partition function in this limit (at $\beta=2$) can be used to analyze the emission of highly energetic strings from a decaying D-brane. This is reminiscent of
\cite{Gross:1987kza}, where Gross and Mende used electrostatic equilibrium
conditions to identify the dominant saddle-point contributions to string scattering amplitudes at arbitrary
order to determine their high-energy behavior.

The system is analogous to the circular unitary ensemble (CUE)
of random matrix theory (an early observation of this is \cite{Larsen:2002wc}), where the connection to the Coulomb gas is a standard tool.
Our method indeed resembles the electrostatic derivation of Wigner's semi-circle
law \cite{wigner} (see also \cite{Forrester}).
The partition function (\ref{eq:Zredef}) is related to an expectation value of a periodic function in the circular unitary ensemble (CUE),
\be \label{eq:CUEexp}
 Z_{n}(\beta;\{\xi_a\};N)  = \prod_{a<b} |w_a-w_b|^{\beta \xi_a\xi_b}\cdot
 \left\langle \prod^n_{a=1} \prod^N_{i=1} |1-w_ae^{-it_i}|^{\beta \xi_a} \right\rangle_{{\rm CUE}_\beta (N)} \ ,
\ee
where $\beta =1,2,4$ correspond to ensembles of symmetric, unitary, and self-dual quaternion unitary matrices $U(N)/O(N)$, $U(N)$, and $U(2N)/Sp(2N)$,
other values define the general $\beta$-ensemble.\footnote{Random matrix interpretations of the generic circular unitary $\beta$-ensembles have been constructed in \cite{killipnenciu}.}

Let us expand the partition function at large $N$, with $N/\xi_a$ fixed. We anticipate the following series expansion:
\be \label{eq:logZexp}
 \log Z_n (\beta;\{\xi_a\};N) = C_0 N^2 +C_1 N^1 + C_2 N^0 + \cdots \ ,
\ee
where the 
coefficients $C_k$ generally depend on the ratios $N/\xi_a$ and may contain logarithms of $N$.
Using a saddle-point-like approach we divide the exponent into the leading contribution and fluctuations,
\bea \label{saddlept}
 \log Z_{n}(\beta;\{\xi_a\};N) &=& - \beta H_0 + \Delta_f \nn \\ \label{eq:saddle}
                               &=& - \beta \mE + \Delta_d + \Delta_f \ ,
\eea
where $H_0$ is the minimum value of the Hamiltonian \eq{Ham} and $\Delta_f$ arises from ``fluctuations'' around the minimum value of the Hamiltonian.
We further divided $H_0$ into
$\mE$, the (total) electrostatic energy of the minimum configuration approximated with continuous charge densities, and the corrections $\Delta_d$ from the difference of the energies of the discrete and continuous systems.
These corrections are analyzed in detail in Appendix~\ref{app:ooN} and they are shown to contribute only at next-to-leading order [\order{N}].
Hence the leading term $C_0 N^2$ is determined by the first term $- \beta \mE$ in \eq{saddle}. Indeed,
the energy $\mE$ contributes only at \order{N^2}, so
\be\label{lead}
 C_0 N^2 = - \beta \mE \ ,
\ee
and the next-to-leading coefficient $C_1$ arises from the corrections $\Delta$ (see Appendix~\ref{app:ooN}).

\subsection{Calculation of the leading term}

We now discuss how the leading term $C_0 N^2 = - \beta \mE$ is calculated. The first
ingredient is the (complex) potential $V(w)$. The physical potential $U$ is the real part,
\be
 U (w) = \re  V (w) \ .
\ee
The potential is sourced by the continuous charge density $\rho (t)$
(which we use to approximate the $N$ unit charges in the $N\rightarrow \infty$ limit, with the
normalization $N=\oint dt \rho (t)$) and the external
charges $\xi_a$,
\be \label{eq:Vdef}
 V(w) = - \int dt\rho(t) \log\left(w-e^{it}\right) - \sum_a \xi_a \log(w-w_a)  \ .
\ee
We fixed the zero level of the potential by setting the possible constant term to zero. We then need to
find $V(w)$ and $\rho (t)$ in the equilibrium configuration. The continuous density consists only of positive elementary charges.

A characteristic feature of the ensuing electrostatic configurations is the presence of gaps in the charge distribution $\rho(t)$. Since the
distribution is made of equal (positive) unit charges, $\rho(t)$ cannot become negative. Hence, the charge distribution may vanish in the
vicinity of an external charge if it is  close enough to the unit circle.
A detailed discussion of this phenomenon in the case of bulk two-point amplitude can be found in \cite{Jokela:2009fd}.
Thus, we are interested in configurations where $\rho(t)$ is strictly positive within a domain $C$ of the unit circle and vanishes elsewhere; the gaps are created by the repulsive force from the positive external charges $\xi_a$.

The problem of finding the equilibrium configuration becomes a
standard potential problem of classical electrostatics in the presence of external charges $\xi_a$ and a conductor which fills the domain $C$.
In the equilibrium configuration the physical potential
$U(e^{it})$ must be constant within $C$,
\be
 U (e^{it}) = U_0 \ , \ e^{it} \in C  \ .
\ee
If the potential would vary within $C$, the charges on the conductor would be subject to forces, and move until the potential adjusts to a constant value.

The electrostatic energy is defined as the energy of the system in the equilibrium configuration, and may be expressed in terms of $\rho_0(t)$ as
\bea \label{eq:Etotdef}
 \mE &=& -\inv{2}\int dt_1 dt_2 \rho_0(t_1) \rho_0(t_2) \log\left|e^{it_1}-e^{it_2}\right| \nn\\
  && - \sum_{a=1}^n \xi_a \int dt\rho_0(t) \log\left|e^{it}-w_a\right|- \sum_{1\le a<b \le n} \xi_a \xi_b \log|w_a-w_b| \nn \\
  &\equiv& \mE_{c} + \mE_{c-\xi}+ \mE_{\xi} \ .
\eea
We again set a possible constant contribution to zero so that the energy matches with the continuum limit of the Hamiltonian \eq{Ham}.
By using the fact that the physical potential $U(e^{it})$ is constant within the domain $C$ of nonzero $\rho_0(t)$, we can express the result as
\be \label{eq:Etotres}
 \mE  = \frac{N}{2} U_0 + \inv{2}\mE_{c-\xi}+ \mE_{\xi} \ ,
\ee
where the first term includes one half of the ``interaction energy'' $\mE_{c-\xi}$.
Hence, to calculate $\mE$ and the leading behavior of $Z_n$ for large $N \sim \xi_a$, it is sufficient to find $U_0$ and $\mE_{c-\xi}$. We present a general method for this in Section~\ref{app:genmet}. In the rest of this paper we will drop the subscript $0$ from the charge density and the potential, so $\rho(t)$ and $U$ will refer to the quantities at equilibrium.

For a system with several external charges many distinct gaps are possible.
Their locations
are fixed by the requirement that $\rho(t)$ must vanish continuously at their endpoints.\footnote{If the number of conductors $\hat n \ge 2$, the domain of nonzero $\rho(t)$ is disconnected, and the potential differences between the connected sub domains are also free parameters. In this case,  energy is minimized when these differences are all zero.}
We will next present the potential problem and find explicit solutions. The reader who is interested in applications of the
solutions may wish to jump to Section~\ref{sec:1pt} upon first reading.

\subsection{The potential problem and its solution}\label{app:pot}

\begin{figure}[ht]
\includegraphics[width=\textwidth]{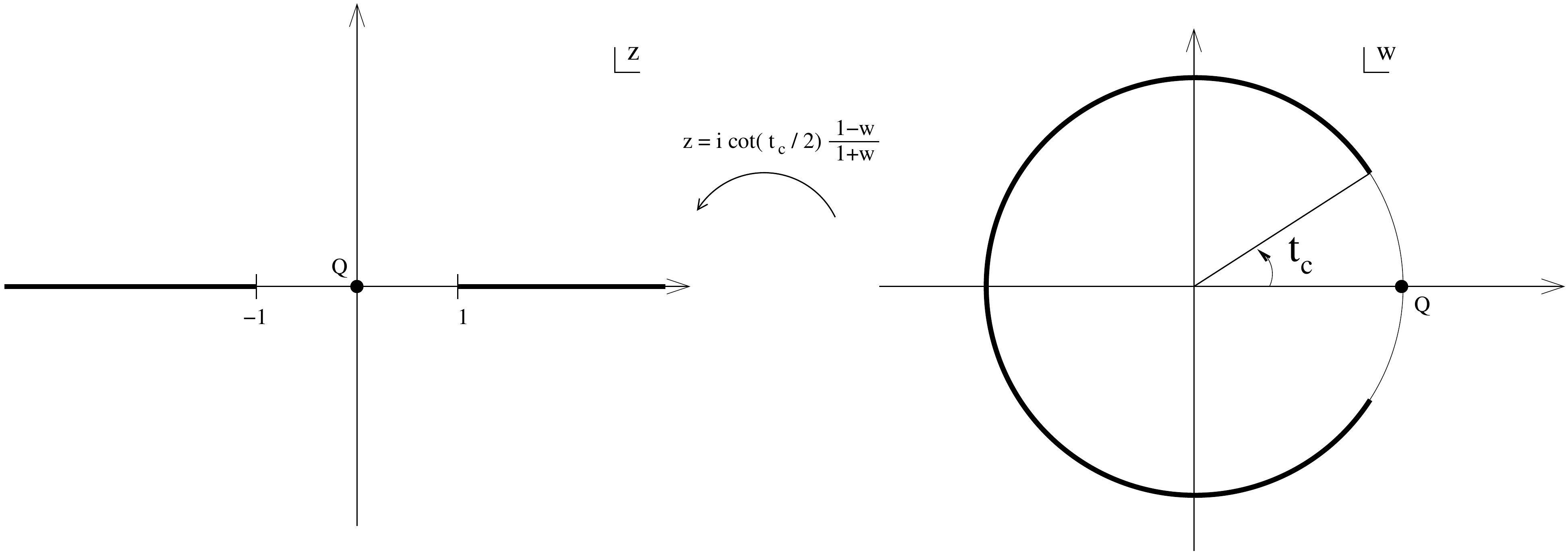}%
\caption{The conformal map that takes the conducting plane to the real axis. A possible mapping of a pointlike charge $Q$ is also illustrated.}
\label{fig:map}
\end{figure}

We are interested in solving potential problems where the conductors are pieces of the arc of the unit circle.
We find it useful to map the disk onto the upper half plane (UHP) by the conformal map (see Fig.~\ref{fig:map})
\be \label{eq:confmap}
 w \mapsto z = i \zeta \frac{1-w}{1+w} \equiv q(w)
\ee
with the understanding that we work in the compactified complex plane with the $\infty$-point included. Here $\zeta$ is a real parameter.

\subsubsection{Electric field construction} \label{app:genmet}

Let us first review a powerful approach which allows us to find compact expressions for electric fields that correspond to configurations with arbitrary numbers of produced strings. The general potential problem has as conductors $\hat n$ separate arcs of the unit circle. It is again useful to map the unit circle on the real axis.
The analytic properties of the complex electric field allow us to write down the solution in a special form \cite{Forrester}. Let us assume for a moment that we have two conductors, $\hat n=2$, which lie at $[a,b]$ and at $[c,d]$. We can then define the following functions
\bea\label{gh}
 g(z) &=& \sqrt{z-a} \sqrt{z-b}\sqrt{z-c}\sqrt{z-d} \nn\\
 h(z) &=& \sqrt{z-a} \sqrt{b-z}\sqrt{c-z}\sqrt{d-z} \ ,
\eea
where the principal branch of the square root function is used.
Then $g(z)$ is analytic everywhere except on the conductors where it has branch cuts, whereas $h(z)$ is analytic over the conductors, but has branch cuts in the gap regions.

Generalization to $\hat n$ conductors is found by adding more terms to $g(z)$ and $h(z)$. We choose the branches such that
\be \label{eq:hgrel}
 h(z) = \mp i g(z) \ ,
\ee
where the minus sign holds in the upper half plane and the plus sign holds in the lower half plane.
It is then possible to show that, when exposed to an appropriate external (conjugate) electric field $E_{\rm ext}(z)$, a solution for the (conjugate) electric field due to the conductors reads
\be
 E_c(z) = \inv{\pi} g(z) \int_\mathrm{cond} dt \frac{\re E_{\rm ext}(t)}{(z-t)h(t)} \ ,
\ee
where the integration is over the conductors. Notice that the solution is bounded at the conductor endpoints.
It is well defined everywhere except on the conductors, where it has discontinuities. The charge density
is proportional to the discontinuity and becomes
\be
 \rho(z) =  \inv{\pi^2}h(z) \int_\mathrm{cond} dt \inv{h(t)} \frac{\re E_{\rm ext}(z)-\re E_{\rm ext}(t)}{z-t} \ .
\ee

The $2\hat n$ endpoints of the conductors are not arbitrary parameters but are subject to constraints.
In particular, these parameters (the constants $a$, $b$, $c$, and $d$ in the $\hat n=2$ case) are partially fixed by the asymptotics of the electric field
\be
 E_c(z) \sim \frac{N}{z}
\ee
for $z \to \infty$ where $N$ is the total electric charge on the conductors. This gives
\bea \label{eq:constraints}
0 &=& \int_\mathrm{cond} dt \frac{t^k\re E_{\rm ext}(t)}{h(t)} \ ,\qquad k=0, \ldots ,\hat n-1 \nn\\
\pi N &=& \int_\mathrm{cond} dt \frac{t^{\hat n}\ \re E_{\rm ext}(t)}{h(t)} \ .
\eea
There are $\hat n-1$ parameters left free, which are identified as the potential differences between the conducting planes. The minimum energy is reached when the potential differences are equal to zero. Therefore, we require
\be
 \re \int_{\mC_k} dz E(z) = 0\ , \qquad k=2, \ldots ,\hat n  \ ,
\ee
where
\be
 E(z) = E_c(z)+ E_{\rm ext}(z)
\ee
is the total electric field and the curve $\mC_k$ connects the first conductor with the $k$th one without intersecting the other conductors, for example.

For the case of interest to us, the external field is due to $n$ point-like charges,
\be
 E_{\rm ext}(z) = \sum_{a=1}^n \xi_a \inv{z-z_a} \ .
\ee
For this configuration the integrals above can be calculated exactly by using contour integration methods. We find
\bea \label{eq:Esol}
 E_c(z) & = & \frac{g(z)}{2}\sum_a\xi_a \left[\inv{g(z_a)(z-z_a)}+\inv{g(z_a^*)(z-z_a^*)}\right]-\sum_a\xi_a \frac{z-x_a}{(z-z_a)(z-z_a^*)} \nn\\
 E(z)   & = &  \frac{g(z)}{2}\sum_a\xi_a \left[\inv{g(z_a)(z-z_a)}+\inv{g(z_a^*)(z-z_a^*)}\right]+\sum_a\xi_a \frac{iy_a}{(z-z_a)(z-z_a^*)} \nn\\
\rho(z) & = & -\frac{h(z)}{2\pi}\sum_a\xi_a \left[\inv{g(z_a)(z-z_a)}+\inv{g(z_a^*)(z-z_a^*)}\right] \ ,
\eea
where $z_a=x_a+iy_a$. The asymptotic conditions imply
\bea \label{eq:constraints2}
 0             & = & \sum_a\xi_a \left[\frac{z_a^k}{g(z_a)}+\frac{\left(z_a^*\right)^k}{g(z_a^*)}\right], \qquad k=0, \ldots, \hat n-1 \nn\\
 N+\sum_a\xi_a & = & \inv{2}\sum_a\xi_a \left[\frac{z_a^{\hat n}}{g(z_a)}+\frac{(z_a^*)^{\hat n}}{g(z_a^*)}\right] \ .
\eea

The potential and the total energy of the system are of special interest to us. We can integrate the complex electric field in \eq{Esol} to give
\bea \label{eq:Vsol}
 V_c(z) & = &  V(z) + \sum_a \xi_a \log(z-z_a) \nn\\
 V(z)   & = & \inv{2}\sum_a\xi_a\left[-\log(z-z_a)+\log(z-z_a^*)\right] \nn\\
        &   & +\lim_{R \to \infty} \Bigg\{\inv{2}\int_z^R dt g(t)\sum_a\xi_a  \left[\inv{g(z_a)(t-z_a)}+\inv{g(z_a^*)(t-z_a^*)}\right]\nn\\
        &   & - \left(N+\sum_a\xi_a\right)\log R\Bigg\} \ .
\eea
Notice that this solution has the asymptotics required by  our definition \eq{Vdef} (or more precisely by its natural generalization to the half plane)
\beq
  V(z)  =  - \left(N+\sum_a\xi_a\right)\log z + \morder{\frac{1}{z}} \ ,
\eeq
so that the constant term of the expansion in $1/z$ is zero.

Let us then write down explicitly the solutions for the disk, \ie, apply the transformation $q(w)$ of \eq{confmap} (with, say $\beta=1$). To do so, we should start with the configuration where charges $\xi_1,\ldots, \xi_n$ are located at arbitrary points $z_1 = q(w_1),\ldots , z_n= q(w_n)$, and an additional charge $\xi_{n+1}=-N - \sum_a \xi_a$ is located at $z=q(\infty)$. Let us denote the corresponding electric field and potential, as given by eqs.~\eq{Esol} and~\eq{Vsol}, by $\tilde E$ and $\tilde V$, respectively.

The electric field on the disk may be then written as
\beqa \label{eq:Emap}
 E(w)   &=& \tilde E(q(w)) \nn\\
 E_c(w) &=& \tilde E_c(q(w)) - \frac{N}{q(w)-q(\infty)}
\eeqa
and integration gives the potential on the disk
\beqa \label{eq:VVcdef}
 V(w)   &=& \tilde V(q(w)) - \lim_{R \to \infty}\left[\int_{q(R)}^\infty dz \tilde E(z) +\left(N+\sum_a\xi_a\right)\log R\right] \nn\\
 V_c(w) &=& \tilde V_c(q(w)) + N \log \left[q(w)-q(\infty)\right]  \nn\\
        &&- \lim_{R \to \infty}\left[\int_{q(\infty)}^R dz \tilde E(z) -N \log R\right] - N\log \left.\frac{d}{dt}\, q\left(1/t \right)\right|_{t=0}  \ .
\eeqa
Here the integrals are independent of $w$. They were added to achieve the correct asymptotics,
\beq
 V(w) =  - \left(N+\sum_a\xi_a\right)\log w + \morder{\frac{1}{w}} \ ; \qquad V_c(w) =  - N \log w + \morder{\frac{1}{w}} \ .
\eeq

Finally, the potential on the conductors is found as
\beq \label{eq:Udef}
 U =\re\, V(w_0)\ ,
\eeq
where $w_0$ is any point on the conductors, and the ``interaction'' energy is given by
\beq \label{eq:Ecxidef}
 \mE_{c-\xi} = \sum_a \xi_a \re \, V_c(w_a) \ .
\eeq
The total energy then reads (see \eq{Etotres} above)
\beq
 \mE = \frac{N}{2}U + \inv{2}\mE_{c-\xi} + \mE_{\xi} \ .
\eeq

\subsubsection{Mapping approach} \label{app:map1pt}

An alternative and complementary approach is found by mapping conformally the complement of the conductors to such a region where the potential solution is known.

In the case of one conductor, say at  $\arg w = t_c \ldots 2\pi-t_c$, we may map the conductor to pieces of the real axis $]-\infty,-a]\cup [a,\infty[$, by choosing $\zeta=a\cot(t_c/2)$ in \eq{confmap} (see Fig.~\ref{fig:map}). For our purposes it is sufficient to solve for the Green function for this configuration, \ie, the solution for one external charge at an arbitrary point.

At this point we could use the formulas of the previous subsection to directly construct the electric field and the potential. However, we may also apply a further transformation
\beq
 z \mapsto i \sqrt{\frac{a+z}{a-z}}
\eeq
that maps the complex plane to the upper half plane and the conductor as the real axis. The Green function is then found by inserting a mirror charge on the lower half plane and mapping back. The obtained solution reads
(for an external charge $Q$ at $z=z_Q$)
\beq \label{eq:V1pt}
 V(z) = -Q\log\frac{\sqrt{\frac{a+z}{a-z}}-\sqrt{\frac{a+z_Q}{a-z_Q}}}{\sqrt{\frac{a+z}{a-z}}+\sqrt{\frac{a+z_Q^*}{a-z_Q^*}}} \ ,
\eeq
where the physical potential is the real part.

The solution of the potential problem of a special interest to us (with the conductor at real axis)
is the sum of \eq{V1pt} with $z_Q=0, Q=\xi$ and $z_Q=-i\zeta, Q=-N-\xi$, where $-i\zeta=-i a \cot\left(t_c/2\right)$ is the image of $w=\infty$ in the conformal map \eq{confmap}:
\beq\label{eq:realV}
  \tilde V(z) = -\xi \log\frac{\sqrt{\frac{a+z}{a-z}}-1}{\sqrt{\frac{a+z}{a-z}}+1}+ (N+\xi)\log\frac{\sqrt{\frac{a+z}{a-z}}-\sqrt{\frac{a-i\zeta}{a+i\zeta}}}{\sqrt{\frac{a+z}{a-z}}+\sqrt{\frac{a+i\zeta}{a-i\zeta}}} +V_0 \ .
\eeq
In (\ref{eq:realV}) the $V_0$ is a constant. The charge density is found as the discontinuity of the electric field $-\tilde V'(z)^*$ on the real axis:
\beqa
 \tilde \rho(z) &=& \frac{|z|}{\pi \sqrt{z^2-a^2}}\left[\frac{(N+\xi)\sqrt{a^2+\zeta^2}}{z^2+\zeta^2}-\frac{\xi a}{z^2}\right] \nn\\
         &=& \frac{\xi\zeta^2\sqrt{z^2-a^2}}{\pi a |z|\left(z^2+\zeta^2\right) }\ , \qquad |z|>a \ ,
\eeqa
where
\beq
 \frac{\zeta}{a} = \frac{\sqrt{N(N+2\xi)}}{\xi}
\eeq
guarantees that the charge density vanishes at $z=\pm a$.
By using the map \eq{confmap}, the complete solution with the conductor on the unit circle can be written as
\bea
 \label{eq:Vcirc}
 V(w) &=& -\xi \log \frac{\sqrt{\frac{we^{it_c}-1}{-w+e^{it_c}}} - 1}{\sqrt{\frac{we^{it_c}-1}{-w+e^{it_c}}} + 1}+(N+\xi) \log \frac{\sqrt{\frac{we^{it_c}-1}{-w+e^{it_c}}} - e^{ i\frac{t_c-\pi}{2}}}{\sqrt{\frac{we^{it_c}-1}{-w+e^{it_c}}} + e^{- i\frac{t_c-\pi}{2}}} +V_0\\
 \label{eq:rhocirc}
\rho(t) &=& \frac{\xi}{2 \pi}\re \sqrt{\cot^2\frac{t_c}{2}-\cot^2\frac{t}{2}} \ ,
\eea
where $\sin(t_c/2) = \xi/(N+\xi)$,  and $t =\arg w$. Notice that the charge density vanishes for $t=-t_c\ldots t_c$ as the square root is purely imaginary.
We use the standard principal branch ($\re (\sqrt{z})>0$ if $|\arg z| < \pi$) for the square root.

\begin{figure}[ht]
\includegraphics[width=\textwidth]{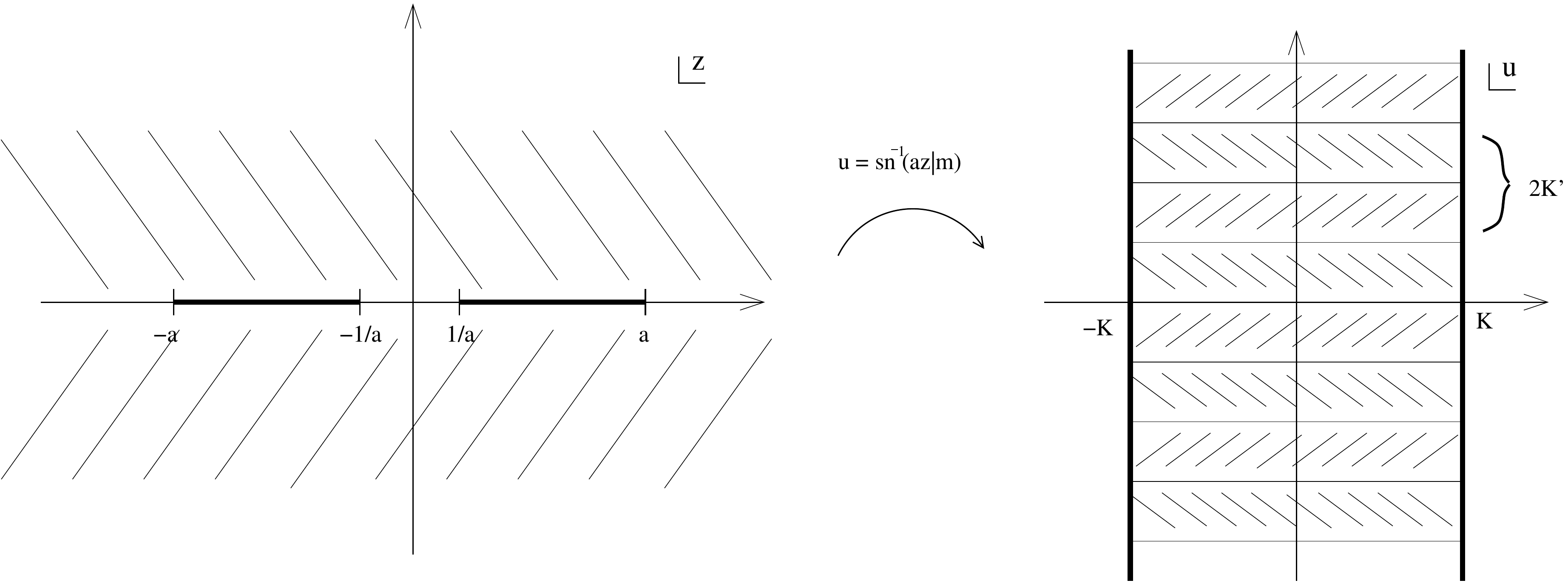}%
\caption{The conformal map that maps the upper and lower half planes to rectangles.}
\label{fig:reflect}
\end{figure}

\subsubsection{Two-gap case}

Let us then go on to the much more complicated case of two conductors.
Without any loss of generality, we may assume that the gaps lie at $]{-1/a},{1/a}[$ and at $]{-\infty},{-a}[\ \cup \ ]{a},{\infty}[$, where $a>1$. Hence we need to solve the potential problem for two conducting planes at $[{-a},{-1/a}]$ and at $[{1/a},{a}]$. Boundary values of the potential could be taken to be  different but we assume that they are equal as this will minimize the energy.

It is useful to further map the problem to a rectangular region. The mapping can be constructed by using the Schwarz-Christoffel formula:
\be
 z \to u = p(z)  \propto \int_0^z \frac{dz'}{\sqrt{(z'+a)(z'+1/a)(z'-1/a)(z'-a)}} \propto \jsn^{-1}(az|m) \ ,
\ee
where $\jsn^{-1}$ is the inverse of the Jacobi elliptic sine function $\jsn$ and $m=a^{-4}$ is the (elliptic) parameter. We fix the normalization to
\be
  p(z) = \jsn^{-1}(az|m) \ .
\ee
Then we find
\bea
 p(0)             & = & 0 \nonumber \\
 p(\pm 1/a)       & = & \pm K(m) \equiv \pm K
\nonumber \\
 p(\pm a + \ieps) & = & \pm K(m) + i K(1-m)   \equiv \pm K + i K'\nonumber \\
 p(i \infty)      & = & i K(1-m) \ ,
\eea
where $K$ is the complete elliptic integral of the first kind.
Hence the upper half plane is mapped on the rectangle $]{-K},{K}[\ \times\ ]{0},{K'}[$ such that the conductors form the vertical edges (see Fig.~\ref{fig:reflect}). Since the gap at  $]{-1/a},{1/a}[$ is mapped to the real axis, we may use the Schwarz reflection principle to argue that the lower half plane is mapped to $]{-K},{K}[\ \times \ ]{-K'},{0}[$ with again conductors on the vertical edges (if the mapping is continued analytically through this gap). Interestingly, if Schwarz reflection principle is used to analytically continue the mapping through the other gap, we see that the lower half plane is also mapped to  $]{-K},{K}[\ \times \ ]{K'},{2K'}[$. Repeating the analytic continuation through the gaps we obtain a conformal mapping from the associated Riemann surface to the strip $]{-K},{K}[\ \times \ \mathbb{R}$, which is periodic in the $u$ plane with the period $2 K' = 2 K(1-m)$ (see Fig.~\ref{fig:reflect}).\footnote{We can also remove the ambiguity of periodicity by mapping the strip to the annulus with the exponential map. We, however, find it more convenient to work on the strip.}

Consequently, we need to look for $2 K'$-periodic solutions in the strip $]{-K},{K}[\ \times \ \mathbb{R}$ with conducting walls. The empty space solution is trivial,
\be
 V(u) = C u \ .
\ee
The solution with a point-like charge $Q$ at $u=u_Q$ (or more precisely at $u=u_Q+2 i K' n$ as required by periodity) is also well known and reads
\be \label{eq:stripGreen}
 V(u) = - Q \log \frac{\theta_1\left(\pi\frac{u-u_Q}{4K}|i\frac{K'}{2K}\right)}{\theta_1\left(\pi\frac{u+u_Q^*-2K}{4K}|i\frac{K'}{2K}\right)} \ ,
\ee
where $\theta_1$ is a Jacobi elliptic theta function. The result equals the field generated by an infinite lattice of positive and negative mirror charges, which are located at the zeros of the numerator and the denominator in \eq{stripGreen}, respectively. Notice that the half-period ratio of the solution $\tau = iK'/2K = iK(1-m)/2K(m)$ is exactly one half of the half-period ratio of the mapping $p$ above.

The above solutions can be used to construct the potential in all cases of interest to us. However, the analytic expressions of most physical quantities are quite cumbersome. In some important special cases there are however drastic simplifications. It is useful to split the potential into symmetric and antisymmetric parts (with respect to the imaginary axis), $V=V_S+V_A$, where
\bea
 V_S(u) & = & -\frac{Q}{2} \log \left[ \frac{\theta_1\left(\pi\frac{u-u_Q}{4K}|i\frac{K'}{2K}\right)}{\theta_1\left(\pi\frac{u+u_Q^*-2K}{4K}|i\frac{K'}{2K}\right)}\frac{\theta_1\left(\pi\frac{u+u_Q^*}{4K}|i\frac{K'}{2K}\right)}{\theta_1\left(\pi\frac{u-u_Q-2K}{4K}|i\frac{K'}{2K}\right)}\right] \\
V_A(u)  & = & -\frac{Q}{2} \log\left[ \frac{\theta_1\left(\pi\frac{u-u_Q}{4K}|i\frac{K'}{2K}\right)}{\theta_1\left(\pi\frac{u+u_Q^*-2K}{4K}|i\frac{K'}{2K}\right)}\frac{\theta_1\left(\pi\frac{u-u_Q-2K}{4K}|i\frac{K'}{2K}\right)}{\theta_1\left(\pi\frac{u+u_Q^*}{4K}|i\frac{K'}{2K}\right)} \right]\ .
\eea
Now the symmetric solution may be written (up to a constant factor) in terms of the elliptic trigonometric functions, and the asymmetric term simplifies also to
\bea
  V_S(u) & = & -\frac{Q}{2} \log\left[ \jsc\!\left(\!\frac{u-u_Q}{2}\bigg|m\!\right) \jsc\!\left(\!\frac{u+u_Q^*}{2}\bigg|m\!\right) \jnd\!\left(\!\frac{u-u_Q}{2}\bigg|m\!\right) \jnd\!\left(\!\frac{u+u_Q^*}{2}\bigg|m\!\right)\right]\nn\\
  V_A(u) & = &  -\frac{Q}{2} \log \frac{\theta_1\left(\pi\frac{u-u_Q}{2K}|i\frac{K'}{K}\right)}
{\theta_1\left(\pi\frac{u+u_Q^*}{2K}|i\frac{K'}{K}\right)} \ ,
\eea
where $\jsc$ and $\jnd$ are Jacobi elliptic trigonometric functions and now all the functions have the same half period ratio as the conformal map $p$. The asymmetric potential vanishes on the imaginary axis as well as on the conductors. By using the trigonometry of the elliptic functions, the symmetric term on the $z$ plane may be expressed in terms of elementary functions
\bea \label{eq:symmsol}
 V_S(z) & = & V_S(u=p(z))|_{u_Q=p(z_Q)} \nonumber\\
        & = & - \frac{Q}{2}\log \Bigg[\left(a z \sqrt{1-\frac{z_Q^2}{a^2}} \sqrt{1-a^2 z_Q^2}-a z_Q \sqrt{1-\frac{z^2}{a^2}} \sqrt{1-a^2 z^2}\right) \nonumber \\
        &   &\times \left(a z \sqrt{1-\frac{z_Q^{*2}}{a^2}} \sqrt{1-a^2 z_Q^{*2}}+a z_Q^* \sqrt{1-\frac{z^2}{a^2}} \sqrt{1-a^2 z^2}\right)  \nonumber \\
        &   & \Bigg/ \Bigg( \sqrt{1-a^2 z^2}\sqrt{1-a^2 z_Q^2}+ \sqrt{1-\frac{z^2}{a^2}} \sqrt{1-\frac{z_Q^2}{a^2}}\nonumber \\
        &   &+ azaz_Q\sqrt{1-\frac{z^2}{a^2}} \sqrt{1-\frac{z_Q^2}{a^2}} +\frac{z}{a}\frac{z_Q}{a}
    \sqrt{1-a^2 z^2} \sqrt{1-a^2 z_Q^2}\Bigg)  \nonumber \\
        &   & \Bigg/ \Bigg(\sqrt{1-a^2 z^2}\sqrt{1-a^2 z_Q^{*2}} +\sqrt{1-\frac{z^2}{a^2}} \sqrt{1-\frac{z_Q^{*2}}{a^2}} \nonumber \\
        &   &- azaz_Q^*\sqrt{1-\frac{z^2}{a^2}} \sqrt{1-\frac{z_Q^{*2}}{a^2}}
   -\frac{z}{a}\frac{z_Q^*}{a}\sqrt{1-a^2 z^2} \sqrt{1-a^2 z_Q^{*2}} \Bigg)\Bigg] \ .
\eea
The (conjugate of the) electric field and the charge density on the conductors are given by
\bea \label{eq:symmrho}
 V_S'(z)  & = &-\frac{Q a}{\sqrt{1-\frac{z^2}{a^2}} \sqrt{1-a^2 z^2}}\left[\jns\left(\jsn^{-1}(z)\!-\!\jsn^{-1}(z_Q)\right)+\jns\left(\jsn^{-1}(z)\!+\!\jsn^{-1}(z_Q^*)\right) \right] \nn \\
\rho_S(z) & = & -\frac{Q|z|}{\pi\sqrt{a^2z^2-1} \sqrt{a^2- z^2}} \re\left(\frac{\sqrt{1-a^2z_Q^2} \sqrt{a^2- z_Q^2}}{z^2-z_Q^2}\right) \ ,
\eea
where the parameter of the Jacobi functions is $m=a^{-4}$ and for the charge density $z$ is real with $1/a<|z|<a$. Unfortunately, we were not able to express the antisymmetric term similarly in terms of elementary functions.

\section{The boundary one-point function} \label{sec:1pt}

Consider the boundary one-point partition function:
\beqa \label{eq:Idef}
 Z_{0+1}(\beta,\xi,N) &=& 
 \inv{N!}\int\left[\prod_{i=1}^{N}\frac{dt_i}{2\pi}|1-e^{it_i}|^{\beta\xi}\right]
 \left[\prod_{1\leq i<j\leq N}|e^{it_i}-e^{it_j}|^{\beta}\right] \ ,
\eeqa
which equals the partition function of Coulomb gas on the circle with an external charge $\xi$ at $w=1$, see, \emph{e.g.}, \cite{npt}.
The subscript $0+1$ denotes one boundary charge ($|w_a|=1$), and zero bulk charges  ($|w_a|<1$).
This case is known to correspond to the circular Jacobi ensemble (see, {\em e.g.}, \cite{Forrester}, Sec. 2.9.6; shift $t_i \rightarrow t_i + \pi$), and the partition function can be evaluated exactly using the Selberg integral formula.
We will discuss that in Section \ref{sec:exact}.
Since this case is in analytical control, it is a good starting point for our scaling limit analysis.

Recall that we consider the limit of large $N$ and $\xi$ with $N/\xi$ fixed,
described by classical electrodynamics.
We assume $\xi>0$ so that the integral \eq{Idef} is surely convergent. Thus the $\xi$-charge repels the unit charges creating a gap in the
charge distribution near $t=0$ where the charge density vanishes. Let us say that the gap is created at $t=-t_c \ldots t_c$. Everywhere
else on the unit circle the distribution is such that the electric potential is constant. Note that consistency requires that the charges
at the edge of the gap feel no force. The charge distribution must therefore vanish continuously at $t=\pm t_c$; otherwise it would not be an equilibrium situation.

It is satisfying to find out that our analysis reproduces a result
of  \cite{BNR},  where a matrix model was constructed for the circular Jacobi
ensemble \eq{Idef}. Ref. \cite{BNR} finds that the spectral measure converges to a
probability  measure supported by
an arc of the unit circle, in a limit which corresponds to $N,\xi
\rightarrow \infty$ with $N/\xi$ fixed. Our calculation can thus be viewed
as an alternative, physics motivated, derivation of that result of
\cite{BNR}, and as a test
of our approach which can be applied to a variety of different cases.

\subsection{Large $N$ calculation}

The electrostatic problem which corresponds to the one-point boundary function was already solved above in Section~\ref{app:map1pt}. To illustrate the use of the alternative, more general method described in Section~\ref{app:genmet}, and to keep the discussion uniform with Section~\ref{sec:npts}, we will anyhow outline how the solution is obtained by using the general method. 

We will first map the disk to upper half plane, using the function $q(w)$ of \eq{confmap} with $\zeta=\cot(t_c/2)$:
\beq
 z \mapsto i \cot\left(\frac{t_c}{2}\right) \frac{1-w}{1+w} = q(w) \ ,
\eeq
see Fig.~\ref{fig:map}. As the conductor is located at $]-\infty,-1] \cup [1,\infty[$, we shall choose the functions $g$ and $h$ of (\ref{gh}) as
\beq
 g(z) = \sqrt{1-z}\sqrt{z+1}\ ; \qquad h(z) = - \sqrt{z-1}\sqrt{z+1} \ .
\ee

It is then straightforward to find the ``tilded'' solutions described in Section~\ref{app:genmet}: We set the charges as
\beqa
 \xi_1 = \xi \quad &\mathrm{at}& \quad z=0 \quad (w=1) \nn\\
 \xi_2 =- N- \xi \quad &\mathrm{at}& \quad z=-i\zeta \quad (w=\infty) \ ,
\eeqa
where the auxiliary 
charge $\xi_2$ is the image under $q(w)$ of the charge at infinity on the $w$ plane.
In the constraints \eq{constraints2} only the former equation (with the index $k=0$) gives a non-trivial condition,
\beq \label{eq:1ptcons}
 \xi = \frac{N+\xi}{\sqrt{1+\zeta^2}} \ ,
\eeq
or equivalently  $\sin(t_c/2) = \xi/(N+\xi)$, whereas the latter one is satisfied automatically because of symmetry. By using this constraint together with Eqs. \eq{Esol} for the above configuration, the electric field and the charge density on the half plane simplify to
\beqa
 \tilde E(z)    &=& \frac{\xi \zeta^2 \sqrt{1-z}\sqrt{z+1}}{z\left(z^2+\zeta^2\right)} - \frac{(N+\xi) i \zeta}{z^2+\zeta^2} \nn\\
 \tilde \rho(z) &=& \frac{\xi \zeta^2 \sqrt{z^2-1}}{\pi |z|\left(z^2+\zeta^2\right)}
\eeqa
which can be mapped to the disk by the map $z=q(w)$ as shown in \eq{Emap}.

One can check that the (rather cumbersome) potential matches with the expressions of Section~\ref{app:map1pt} both on the half plane and on the disk. 
However, the potential on the conductor $U$ of Eq. \eq{Udef} and the ``interaction energy'' $\mE_{c-\xi}$ of Eq. \eq{Ecxidef} reduce to very simple forms when $\zeta$ is eliminated by using \eq{1ptcons}. We find
\beqa \label{eq:1ptcontVc}
  U           &=&  \inv{2}\left[-f(N) - f(N+2\xi)+2f(N+\xi)\right] \nn\\
  \mE_{c-\xi} &=& \xi\left[-f(\xi)+f(2\xi)+f(N+\xi)-f(N+2\xi)\right] \ ,
\eeqa
where
\be
 f(x) = x \log x \ .
\ee
Thus, the total energy reads
\bea \label{eq:1ptEtot} \label{eq:1ptcontEx}
 \mE^{\rm(1pt)} &=&  \frac{N}{2} U + \inv{2}\mE_{c-\xi} \nn\\
               &=& \inv{2}\left[-2F(\xi)+ F(2\xi) - F(N) - F(N+2\xi)+ 2F(N+\xi)\right] \ ,
\eea
where
\be
 F(x) = \inv{2}x f(x) = \inv{2} x^2 \log x \ .
\ee
The resulting leading asymptotics of the partition function for $N \to \infty$ with $N/\xi$ fixed is
\bea \label{eq:1ptZas1}
 \log Z_{0+1}(\beta,\xi,N) &\simeq& -\beta \mE  \nn\\
                &=& \frac{\beta}{2}\left[2F(\xi)- F(2\xi) + F(N) +F(N+2\xi)-2F(N+\xi)\right] \ . 
\eea

The charge density becomes, after mapping to the circle,
\be \label{eq:1ptrho}
 \rho(t) = \frac{\xi}{2\pi}\sqrt{\cot^2\left(\frac{t_c}{2}\right)-\cot^2\left(\frac{t}{2}\right)} \ .
\ee
This result agrees with (5.6) in \cite{BNR}. Their $d$ is equal to our
ratio $\xi/N$, which is real valued, so that their $\xi_d=0$.  The gap
angle $\theta_d$ in \cite{BNR},
\be
  \sin \left( \frac{\theta_d}{2} \right) =  \frac{d}{1+d} = \frac{\xi/N}{
1+(\xi/N)}
\ee
is seen to be equal to our gap angle $t_c$. Finally, our charge density
\eq{1ptrho} can easily be rewritten in the form (5.6) of \cite{BNR}.
The result may also be used to calculate the next-to-leading correction to $\log Z$. We find
\be
 \int dt \rho(t)\log2\pi\rho(t) = \xi \log 2 + \inv{2}f(N)+f(N+\xi)-\inv{2}f(N+2\xi) \ .
\ee
Therefore, using Eq. \eq{ooNcorr} from Appendix~\ref{app:ooN}, the next-to-leading result reads
\bea \label{eq:1ptZas2}
 \log Z_{0+1}(\beta,\xi,N)
                &=& \frac{\beta}{2}\left[2F(\xi)- F(2\xi) + F(N) +F(N+2\xi)-2F(N+\xi)\right] \nn\\ 
  && + \frac{\beta-2}{2}\left[\xi \log 2 + \inv{2}f(N)+f(N+\xi)-\inv{2}f(N+2\xi)\right] \nn\\
  && + \left[\frac{2-\beta}{\beta}+\frac{\beta}{2}\log\frac{\beta}{2}-\log\Gamma\left(\frac{\beta}{2}+1\right)\right] N + \morder{\log N} \ .
\eea

Let us make one more comment which is special to $\beta=2$. At this temperature, retaining the leading term of $\log Z_{0+1}$ (the next-to-leading term vanishes), and further considering the limit $N /\xi \gg 1$ we find
\be \label{eq:largeN}
 Z_{0+1}(\beta,\xi,N) \simeq N^{\xi^2} \ .
\ee
This agrees with the leading term of $\log Z_{0+1}$ in the limit of large $N$ with $\xi$ fixed \cite{Fisher}.
Thus the electrostatic result correctly interpolates between the leading terms in these two limits. We have checked this explicitly only for the one-point amplitude, but the result is true in general. This is evident since the large $N$ result of Eq. \eq{largeN} has a simple electrostatic interpretation: the corresponding energy $-(\xi^2\log N)/2$ is the missing self-energy of the small gap created by the external charge in the otherwise flat charge distribution \cite{npt}.

\subsection{Comparison to the result for arbitrary $N$}\label{sec:exact}

The exact result for the partition function \eq{Idef} for general $N$ and  $\beta$ can be found by using the Selberg integral \cite{Forrester,mehta} and reads
\be \label{eq:1ptres}
  Z_{0+1}(\beta,\xi,N) =
                        \prod^{N-1}_{j=0} \frac{\Gamma ((\beta /2)j+\beta \xi +1)\Gamma ((\beta /2)(j+1)+1)}
                        {[\Gamma ((\beta /2)j +(\beta \xi/2)+1)]^2\Gamma((\beta /2)+1)} \ .
\ee
It is straightforward to check that its expansion at large $N$ and $\xi$ equals Eq.~\eq{1ptZas2}. For $\beta=2$ the result \eq{1ptres} can be expressed in terms of Barnes $G$ functions as
\be \label{eq:1ptatb2}
 Z_{0+1}(2,\xi,N) =  \frac{G(\xi+1)^2}{G(2\xi+1)}\frac{G(N+2\xi+1)G(N+1)}{G(N+\xi+1)^2} \ .
\ee
In particular, either by using the general result of Eq.~\eq{1ptres} of by using the asymptotics of the Barnes $G$ function in Eq.~\eq{1ptatb2}, one can check the \order{N} term in  Eq.~\eq{1ptZas2} indeed vanishes at $\beta=2$.

The ``exact'' counterpart of the continuum charge density \eq{1ptrho} (or \eq{rhocirc}) is the distribution function of unit charges in the Coulomb gas. It
can be acquired by leaving one of the $t_i$'s (say $t_1$) unintegrated in \eq{Idef}. That is,
\be
 \rho_N(t_1) =  \inv{Z(\xi;N) N!} \int\left[\prod_{i=2}^{N}\frac{dt_i}{2\pi}\right]\left[\prod_{i=1}^{N}|1-e^{it_i}|^{2\xi}\right]
 \left[\prod_{1\leq i<j\leq N}|e^{it_i}-e^{it_j}|^{2}\right] \ ,
\ee
where we again fixed $\beta=2$.
In the large $N$ limit, with $\xi/N$ fixed, we expect that
\be \label{eq:denslim}
  \rho_N(t) \  \substack{\phantom{A} \\\longrightarrow \\ {\scriptscriptstyle N \to \infty}} \ \inv{N} \rho(t) \ ,
\ee
where $\rho(t)$ was given in Eq.~\eq{1ptrho}. In fact, $\rho_N(t_1)$ (as well as the higher point distribution functions) is known exactly for $\beta=2$ and arbitrary $N$ (see Chapter~4 of \cite{Forrester}). By using this result, it is possible to check the limit \eq{denslim} explicitly.
Thus the exact distribution functions provide yet another way to derive Eq.~\eq{1ptrho}.
We have also checked the limit by comparing the first few Fourier coefficients of the functions $\rho$ and $\rho_N$, which are calculable for all $\xi$ and $N$ \cite{Schomerus:2008je}.

\subsection{Lattice path interpretation} \label{sec:lpi}

\begin{figure}[ht]
\begin{center}
\includegraphics[width=0.5\textwidth]{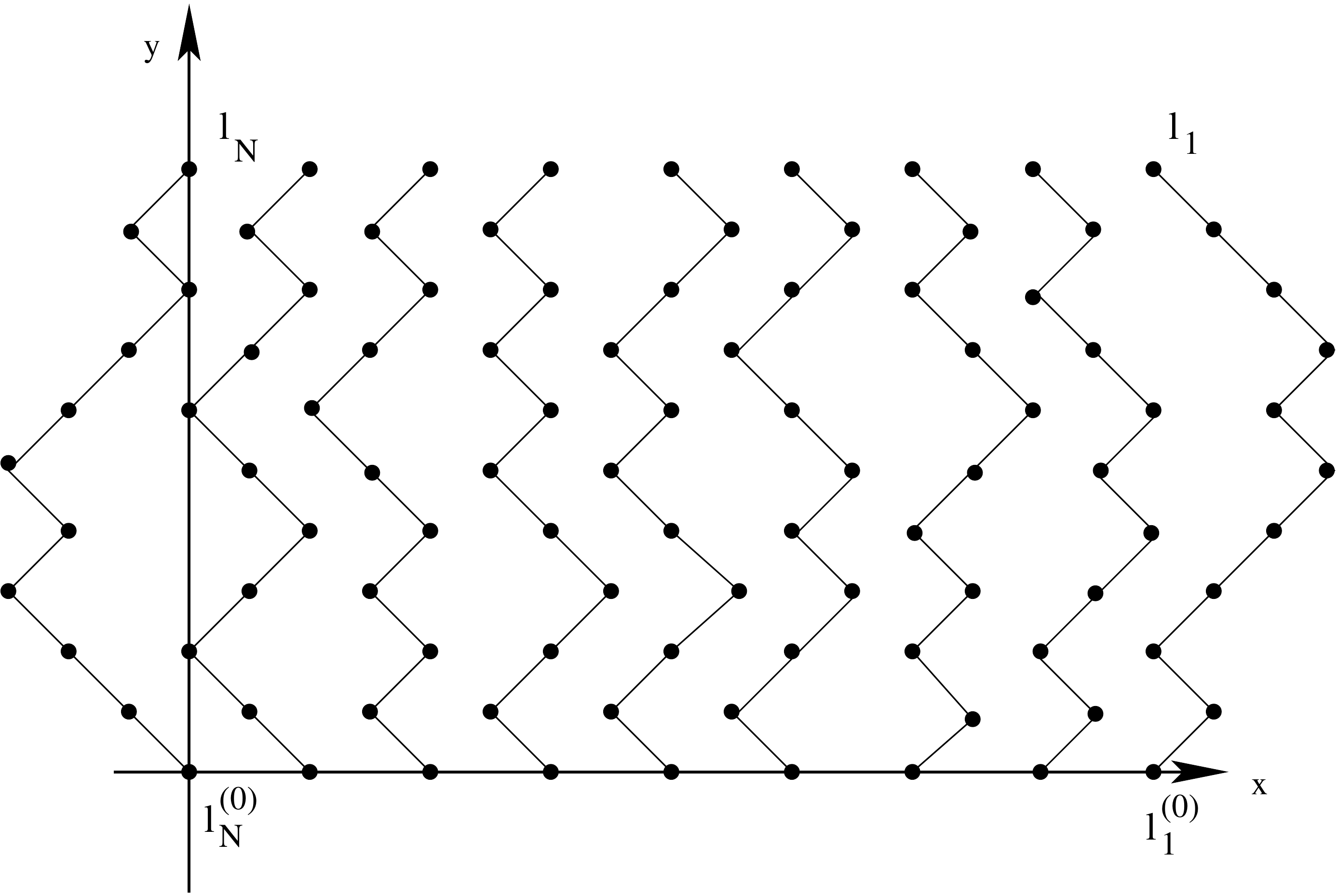}%
\caption{An example of $N=9$ non-intersecting left diagonal / right diagonal lattice paths of
$2n=10$ steps/segments, with endpoints $l_j$ shifted from the starting points $l^{(0)}$ by $r=0$ in the $x$-direction.}
\label{fig:vicwalk}
\end{center}
\end{figure}

We first rewrite the partition function (\ref{eq:Idef}) as a CUE expectation value,
\bea\label{cue-exp}
Z_{0+1}(\beta, \xi ,N)
                       &=& \left\langle \prod^N_{i=1} (1+e^{it_i})^{\beta \xi /2}  (1+e^{-it_i})^{\beta \xi/2}\right\rangle_{{\rm CUE}_{\beta}(N)} \ ,
\eea
where we have shifted $t_i \rightarrow t_i + \pi$. If we set $\beta=2, \xi=n$, we recognize
it as the $n_1=n_2=n$ case of the expectation value appearing in \cite{Forrester} Exercises 8.1,
\be
Z_{0+1}(2, n ,N) = G^{ld/rd}_{2n} (\{ l^{(0)}_j=2(N-j)\}_{j=1,\ldots,N};\{ l_j=2(N-j)+r\}_{j=1,\ldots,N})\Big|_{\stackrel{w^{\mp}_k=1}{k=1,\ldots,2n}}
\ee
with $r=0$. The resulting expression $G^{ld/rd}_n$ denotes the total number of $N$ non-intersecting single
move $ld/rd$ (left diagonal/right diagonal) paths starting from locations $l^{(0)}_j$ and ending at $l_j$ (see Fig.~\ref{fig:vicwalk}) in $2n$ steps (segments), in the lock step model of vicious walkers \cite{Fisherwalk,Forrestwalk}, a variant of the 
vicious walker model introduced to model wetting and melting.


\section{Higher-point boundary functions} \label{sec:npts}


The boundary two-point partition function can in principle be treated in the same manner as the one-point function above. Now we have two charges, $\xi_1$ and $\xi_2$, on the unit circle. When these charges are well distinct they form two separate gaps in the charge distribution of the unit charges at large $N$. However, when the charges approach each other the gaps will merge at some critical separating angle. Hence, we have phases with either one or two gaps in the charge distribution.

The general solution for the two-point boundary function is complicated. Hence we will only discuss two special cases below and restrict to the leading asymptotics. We shall also discuss some simple generalizations to higher-point functions.

\subsection{``Antipodal'' two-point function}

When the two charges are exactly at antipodal points of the circle, the solution has two gaps regardless of the values of $\xi_1$ and $\xi_2$. We can fix the charge $\xi_1$ at $\tau_1=0$ ($w_1 = e^{i\tau_1}=1$) and the charge $\xi_2$ at $\tau_2=\pi$ ($w_2= e^{i\tau_2}=-1$). Then the charge $\xi_1$ ($\xi_2$) is mapped to $z_1=0$ ($z_2=\infty$) in the mapping $z=q(w)$ of Eq. \eq{confmap}. For a proper choice of the parameter $\zeta$ (to be solved below) the configuration is that of Section~\ref{app:map1pt}. In particular, the system is reflection symmetric with respect to the imaginary axis. Hence, the potential on the half plane can be expressed in terms of the symmetric solution of Eq.~\eq{symmsol} as
\be
 \tilde V(z) = \left.V_S(z)\right|_{Q=\xi_1,\ z_Q=0}+\left.V_S(z)\right|_{Q=\xi_2,\ z_Q=i \infty}+\left.V_S(z)\right|_{Q=-(N+\xi_1+\xi_2),\ z_Q=-i\zeta} + V_0 \ ,
\ee
where $V_0$ is a constant.
The charge density is constructed by using Eq.~\eq{symmrho} and becomes
\be
 \tilde \rho(z) = \frac{|z|}{\pi\sqrt{a^2z^2-1}\sqrt{a^2-z^2}}\left[ \frac{(N+\xi_1+\xi_2)\sqrt{1+a^2\zeta^2}\sqrt{a^2+\zeta^2}}{z^2+\zeta^2}-\frac{\xi_1 a}{z^2} -\xi_2 a\right] \ .
\ee
Requiring it to vanish at $z=\pm a$ and at $z=\pm1/a$ fixes
\bea
 \zeta &=& \sqrt{\frac{\xi_2}{\xi_1}}\nn\\
 a &=&\frac{\sqrt{(N+\xi_1)(N+\xi_2)+\xi_1\xi_2 +\sqrt{N(N+2\xi_1)(N+2\xi_2)(N+2\xi_1+2\xi_2)}}}{\sqrt{2\xi_1\xi_2}} \ .
\eea
If we further fix the constant $V_0$ by requiring that
\be
 \tilde V(z) =\xi_2\log z+ \morder{1/z}
\ee
as $z \to \infty$, the solution equals the $\tilde V$ defined in Section~\ref{app:genmet}. It may therefore be mapped to the disk by using Eqs.~\eq{VVcdef}. After a straightforward calculation, the potential and the interaction energies (see Eqs.~\eq{Udef} and \eq{Ecxidef}) become
\bea
 U &=& f(N+\xi_1+\xi_2)-\inv{4}\left[f(N)+f(N+2\xi_1)+f(N+2\xi_2)+f(N+2\xi_1+2\xi_2)\right] \nn\\
 \mE_{c-\xi} &=& (\xi_1+\xi_2)\left[f(N+\xi_1+\xi_2)-\inv{2}f(N+\xi_1+\xi_2)\right]-\frac{\xi_1}{2}f(N+2\xi_1)\nn\\
 &&-\frac{\xi_2}{2}f(N+2\xi_2)- \inv{2} F(\xi_1+\xi_2) + \inv{8}\left[ F(2 \xi_1+ 2 \xi_2)+F(2 \xi_1)+ F(2 \xi_2)\right] \nn\\
 \mE_\xi &=& -\xi_1\xi_2 \log 2 \ ,
\eea
where again $f(x)=x\log x$ and $F(x) = (x^2\log x)/2$. Thus, the leading asymptotics of the partition function for the antipodal configuration is
\bea \label{eq:antipe}
 \log Z_{0+2} & \simeq & -\beta \mE^{\rm(2pt, a)}  = -\frac{\beta}{2} N U -\frac{\beta}{2} \mE_{c-\xi} -\beta \mE_\xi \nn\\
 & = & \beta\bigg\{-F(N+\xi_1+\xi_2)+ F(\xi_1+\xi_2) \nn\\
  &  & + \inv{4}\left[F(N+2 \xi_1+2\xi_2) +F(N+2 \xi_1) +F(N+2 \xi_2) +F(N)\right] \nn \\
       &   & - \inv{4}\left[ F(2 \xi_1+ 2 \xi_2)+F(2 \xi_1)+ F(2 \xi_2)\right] +\xi_1\xi_2\log 2 \bigg\} \ .
\eea

\subsection{Two-point function with equal charges}

Let us then discuss a slightly more complicated special case where the two charges are equal $\xi_1=\xi_2\equiv \xi$ but their locations are arbitrary.
When the angle $\tau \equiv \tau_2-\tau_1$ between the charges is smaller than some critical angle $\tau_c$, there is only one gap.
By rotational symmetry the two charges can be taken to lie at $e^{\pm i \tau/2}$ on the unit circle. Then the mapping \eq{confmap} with $\zeta = \cot\tau/4$ maps the $\xi$-charges to $z=\pm 1$ and the $w=\infty$ point to $z=-i\zeta = -i \cot(\tau/4)$. We can then use the methods of Section~\ref{app:genmet} with
\be
 g(z) = \sqrt{z-b}\sqrt{z+b} \ ,
\ee
where $1>b>0$ and we used the fact that the solution must be symmetric with respect to the imaginary axis. The constraint equation for the parameter $b$ simplifies to
\be \label{eq:1hcons}
 \frac{2\xi}{\zeta\sqrt{1-b^2}} =  \frac{N+2\xi}{\sqrt{b^2+\zeta^2}} \ .
\ee
Following the procedure described above and in Section~\ref{app:genmet}, and after a lengthy calculation, the energy can be expressed as 
\bea
 \mE^{\rm(1g)} &=& - \inv{2}F(N) + F(N+2\xi)-\inv{2}F(N+4\xi) \nn\\
  && + \inv{4}N(N+2\xi)\log\left(\zeta ^2+1\right) +6 \xi^2 \log 2 - \xi^2\log \left(2\sin\frac{\tau}{2}\right) \ .
\eea

For $\tau > \tau_c$ there are two gaps. Then we can use the same choice for the mapping $z=q(w)$ to the half plane as in the one gap phase, with $\zeta = \cot\frac{\tau}{4}$. Now we choose in Section~\ref{app:genmet}
\beq
  g(z) = \sqrt{z+a}\sqrt{z-a}\sqrt{z-b}\sqrt{-z-b} \ ,
\eeq
where we used symmetry to fix two of the parameters defining the end point locations and $a>1>b>0$. The remaining parameters $a,b$ are fixed from the equations \eq{constraints} which simplify to
\bea \label{eq:2ptconstr}
 &\left(N+2\xi\right)\zeta \sqrt{a^2-1}\sqrt{1-b^2} = 2 \xi \sqrt{\zeta^2+a^2} \sqrt{\zeta^2+b^2}& \nn\\
 &\zeta^2\left(a^2-1\right) \Pi\left(1-b^2\big|1-b^2/a^2\right) +\left(a^2+\zeta^2\right) \re \Pi\left(1+b^2/\zeta^2\big|1-b^2/a^2\right) = 0&
\eea
where $\Pi(n|m)$ is the complete elliptic integral of the third kind. Taking the real part of $\Pi$ in the latter equation removes the branch cut ambiguity, which appears since the argument $n=1+b^2/\zeta^2$ of $\Pi(n|m)$ is larger than one.

The critical angle $\tau_c$ may be extracted from \eq{2ptconstr}. For $\tau=\tau_c$ either of the conductors should disappear and the gaps should merge together, \ie, we need to study the limit $b \to 0$ or $a \to \infty$, which yields
\bea
 &&(N+2 \xi ) \sinh ^{-1}\left(\sqrt{\frac{(N+2 \xi )^2 \tan ^2\left(\frac{\tau_c}{4}\right)+4 \xi ^2}{N
   (N+4 \xi )}}\right) \nn\\ &=&{2 \xi }\tanh ^{-1}\left(\frac{2 \sqrt{2} \xi }{\sqrt{N^2-N (N+4 \xi ) \cos
   \left(\frac{\tau_c}{2}\right)+4 N \xi +8 \xi ^2}}\right) \ .
\eea

After a tedious calculation by using the formulas in Section~\ref{app:genmet}, the energy can be expressed as
\be
 \mE^{\rm(2g)} = \frac{2 \left(\zeta ^2+1\right) \xi ^2}{\sqrt{1-b^2}} J_1 + \frac{\left(\zeta ^2+1\right) \xi  (N+2 \xi )}{\sqrt{1-b^2}} J_2 + \mE_{\rm log} +\mE_\xi \ ,
\ee
where
\bea \label{eq:2ptres}
 J_1 &=& \lim_{\eps\downarrow0}\left[ \int_{1+\eps}^a dz \frac{\sqrt{a^2-z^2}\sqrt{z^2-b^2}}{\left(z^2-1\right)\left(z^2+\zeta^2\right)} + \frac{\sqrt{a^2-1}\sqrt{1-b^2}}{2\left(1+\zeta^2\right)}\log\eps  \right] \nn\\
&=&-\frac{\sqrt{a^2-1} \sqrt{1-b^2} \log\! \left(\!\frac{a^2 \left(b^2-2\right)+1}{2 \left(a^2-1\right) \left(b^2-1\right)}\!\right)}{2 \left(\zeta
   ^2+1\right)}-\frac{\left(b^2+\zeta ^2\right) F\!\left(\sin ^{-1}\!\left(\!\sqrt{\frac{a^2-1}{a^2-b^2}}\right)|1-\frac{b^2}{a^2}\right)}{a \left(\zeta ^2+1\right)} \nn\\
&& + \frac{\left(b^2+\zeta ^2\right) \Pi \left(\frac{a^2-b^2}{a^2+\zeta ^2};\sin
   ^{-1}\left(\sqrt{\frac{a^2-1}{a^2-b^2}}\right)|1-\frac{b^2}{a^2}\right)}{a \left(\zeta ^2+1\right)}\nn\\
&& -\frac{\left(1-b^2\right) \Pi \left(1-\frac{1}{a^2};\sin ^{-1}\left(\sqrt{\frac{a^2-1}{a^2-b^2}}\right)|1-\frac{b^2}{a^2}\right)}{a \left(\zeta
   ^2+1\right)}
\nn\\
 J_2 &=& \lim_{\eps\downarrow0}\left[  \int_{\zeta+\eps}^\infty dy \frac{\sqrt{y^2+a^2}\sqrt{y^2+b^2}}{\left(y^2+1\right)\left(y^2-\zeta^2\right)} +\frac{\sqrt{\zeta^2+a^2}\sqrt{\zeta^2+b^2}}{2\zeta\left(\zeta^2+1\right)}\log\eps \right] \nn\\
&=& -\frac{\sqrt{a^2+\zeta ^2} \sqrt{b^2+\zeta ^2} \log \left(\frac{a^2 b^2+2 a^2 \zeta ^2+\zeta ^4}{2 \zeta  \left(a^2+\zeta ^2\right)
   \left(b^2+\zeta ^2\right)}\right)}{2 \zeta  \left(\zeta ^2+1\right)} +\frac{\left(b^2+\zeta ^2\right) F\left(\cot ^{-1}\left(\frac{\zeta }{a}\right)|1-\frac{b^2}{a^2}\right)}{a \left(\zeta ^2+1\right)} \nn\\
&& + \frac{\left(1-b^2\right) \Pi \left(1-\frac{1}{a^2};\cot ^{-1}\left(\frac{\zeta }{a}\right)|1-\frac{b^2}{a^2}\right)}{a \left(\zeta ^2+1\right)}\nn\\
&& -\frac{\left(b^2+\zeta ^2\right) \Pi \left(\frac{(a-b) (a+b)}{a^2+\zeta ^2};\cot ^{-1}\left(\frac{\zeta }{a}\right)|1-\frac{b^2}{a^2}\right)}{a
   \left(\zeta ^2+1\right)}
\nn\\
\mE_{\rm log} &=& 2 \xi ^2 \log \left(\frac{2 \sqrt{2} \zeta }{\zeta
   ^2+1}\right)-\frac{1}{4} (N+2 \xi )^2 \log (2 \zeta ) \nn\\
   \mE_\xi &=& -\xi^2\log \left(2\sin\frac{\tau}{2}\right) \ .
\eea
The calculation was done by using integral tables and checked by Mathematica.
$F(u|n)$ [$\Pi(n;u|m)$] is the incomplete elliptic integral of the first [third] kind.\footnote{The result could possibly be simplified a lot, if the constraint equations could be solved and substituted into \eq{2ptres}. Such simplifications take place for simpler configurations. }

\begin{figure}[ht]
\includegraphics[width=0.5\textwidth]{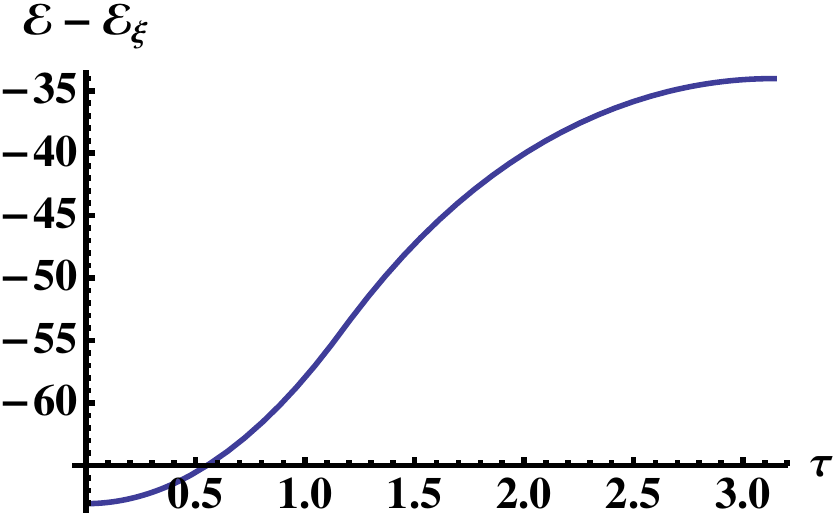}\includegraphics[width=0.5\textwidth]{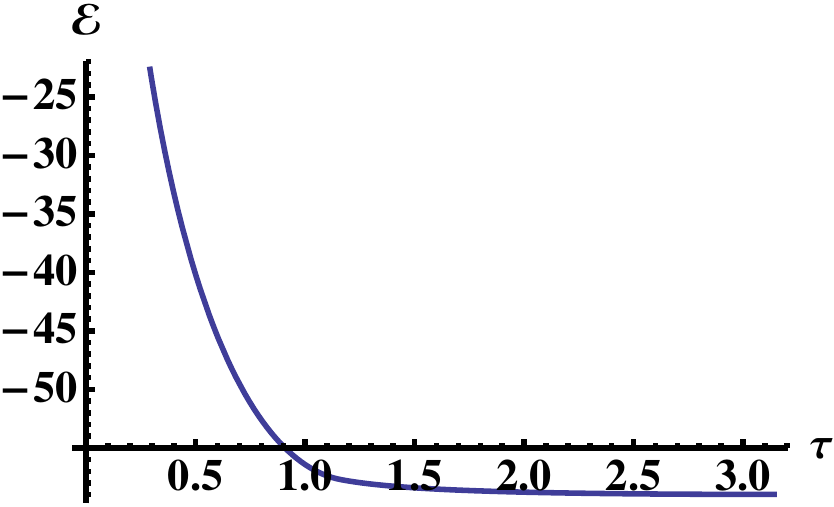}%
\caption{The boundary two-point total energy as a function of $\tau$. We used $\xi=6$ and $N=16$. In the left hand plot the energy of the mutual interaction of the pointlike charges $\mE_\xi$ was excluded. The critical angle is given by $\tau_c\simeq 1.157$.}
\label{fig:Energy}
\end{figure}

The one and two-gap solutions are combined to give the total energy as
\be
 \mE^{\rm(2pt)} = \mE^{\rm(1g)}\theta\left(\tau_c-\tau\right)+\mE^{\rm(2g)}\theta\left(\tau-\tau_c\right)
\ee
in Fig.~\ref{fig:Energy}. Notice that the solutions join smoothly at  $\tau=\tau_c$. When the interaction term $\mE_\xi$ is included, the $\tau$ dependence is relatively flat for large separations ($\tau$ close to $\pi$).

\begin{figure}[ht]
\includegraphics[width=0.5\textwidth]{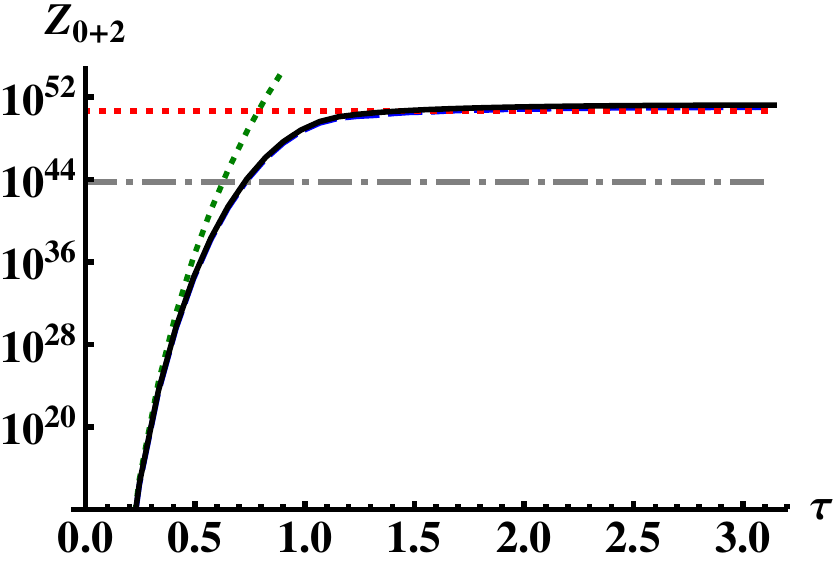}\includegraphics[width=0.5\textwidth]{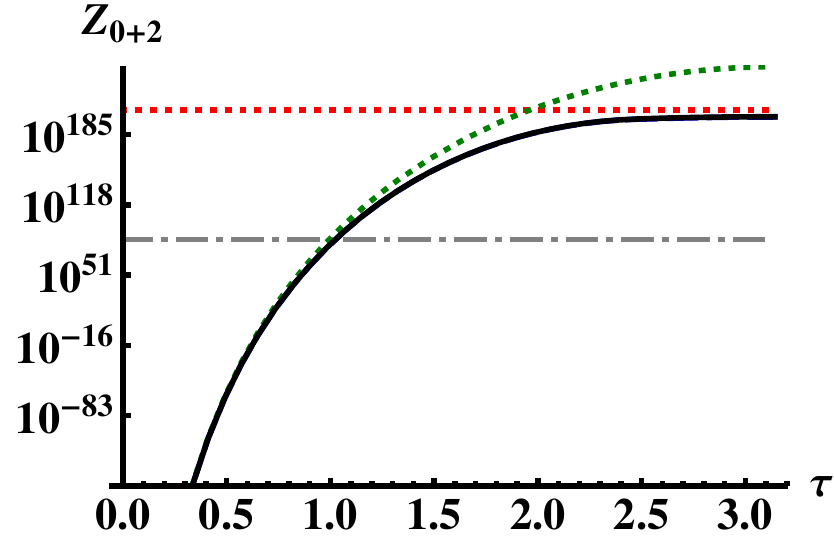}%
\caption{Comparison of the electrostatic large $\xi,N$ approximation of $Z_{0+2}$ to asymptotics of the Toeplitz determinant and to the exact partition function at $\beta=2$ as a function of $\tau$. We used $\xi=6$ and $N=16$ ($N=16$ and $N=\xi$) on the left (right) hand plot. The solid black curve is the leading term obtained by electrostatics given in Eq.~\eq{Zes}. The exact result is shown by the dashed blue curve which is hardly visible since it overlaps with the leading term. The gray dash-dotted horizontal line arises from the asymptotics of the Toeplitz determinant, Eq.~\eq{Zas}. The horizontal red dotted and the green dotted lines are the improved asymptotics of Eq.~\eq{Zimpr} and the small $\tau$ approximation of Eq.~\eq{Ztau}.
}
\label{fig:taudep}
\end{figure}

Let us then compare the leading term of the partition function (at $\xi_1=\xi_2$ and $\beta=2$),
\be \label{eq:Zes}
 Z_{0+2}(\xi,N) \simeq \exp\left(-2\mE^{\rm (2pt)}\right)
\ee
to the exact result and earlier approximations. The exact partition may be calculated at small $N$ by noticing that the partition function at $\beta=2$ defines the Toeplitz determinant (see \cite{VEPA})
\be \label{eq:Zexc}
 Z_{0+2}(\xi_1,\xi_2;N) =  \left|e^{i\tau_1}-e^{i\tau_2}\right|^{2\xi_1\xi_2}\det T_N(h)
\ee
for the function
\be
 h(t) = \left|e^{it}-e^{i\tau_1}\right|^{2\xi_1} \left|e^{it}-e^{i\tau_2}\right|^{2\xi_2} \ .
\ee
The partition function can be evaluated in the limit $N \to \infty$ with $\xi$ fixed by using the asymptotics of Toeplitz determinants \cite{Fisher}. Leading term given in the literature can be written in terms of the Barnes $G$ function as
\be \label{eq:Zas}
  Z_{0+2}(\xi_1,\xi_2;N) \simeq \prod_{a=1}^2 \frac{G(\xi_a+1)^2}{G(2\xi_a+1)}  \frac{G(N+2\xi_a+1)G(N+1)}{G(N+\xi_a+1)^2} \ ,
\ee
when applied to our case. Based on physical considerations we suggested \cite{npt} the improved formula
\bea \label{eq:Zimpr}
  Z_{0+2}(\xi_1,\xi_2;N) &\simeq& \frac{\Gamma(N+\xi_1+1)\Gamma(N+\xi_2+1)}{\Gamma(N+1)\Gamma(N+\xi_1+\xi_2+1)G(N+\xi_1+\xi_2+1)^4} \nn\\
&&\times  \prod_{a=1}^2 \frac{G(\xi_a+1)^2}{G(2\xi_a+1)} G(N+\xi_a+\sum_b\xi_b+1)G(N+\xi_a+1) \ .
\eea
Notice that both above approximations are independent of $\tau$. They are expected to work well when the charges are close to antipodal even for relatively large $\xi$, but they will break down for $\tau \lesssim \xi/N$ (see the discussion in \cite{npt}), as we shall see. In the region of small $\tau$ we may in fact approximate the partition function by
separating the interactions of the external charges and then joining the charges into a single one:
\bea \label{eq:Ztau}
  Z_{0+2}(\xi_1,\xi_2;N) &\simeq& \left|e^{i\tau_1}-e^{i\tau_2}\right|^{2\xi_1\xi_2} Z_{0+1}(\xi_1+\xi_2,N)\\ \nn
  &=&\left|e^{i\tau_1}-e^{i\tau_2}\right|^{2\xi_1\xi_2} \frac{G(\sum_a\xi_a+1)^2}{G(2\sum_a\xi_a+1)} \frac{G(N+2\sum_a\xi_a+1)G(N+1)}{G(N+\sum_a\xi_a+1)^2} \ ,
\eea
where we used the exact result for the one-point function on the second line (or equivalently either of the approximation schemes above, since they all give the same result for the one-point function).

All these approximations are compared in Fig.~\ref{fig:taudep}. As expected, the asymptotic formulas obtained from the asymptotics of Toeplitz determinants work well for the region of large $\tau$ where the partition function is almost constant. Actually, the improved asymptotic formula \eq{Zimpr} (dotted red line) gives a much better approximation than the standard leading term of the Toeplitz asymptotics \eq{Zas} (dash-dotted gray line). The leading term at large $N$ \eq{Zes} (black curve) interpolates between the other approximations, and follows closely the exact result (blue dashed curve), which is obtained numerically from Eq.~\eq{Zexc}.

\subsection{Symmetric $n$-point functions} \label{sec:symm}

\begin{figure}[ht]
\includegraphics[width=0.5\textwidth]{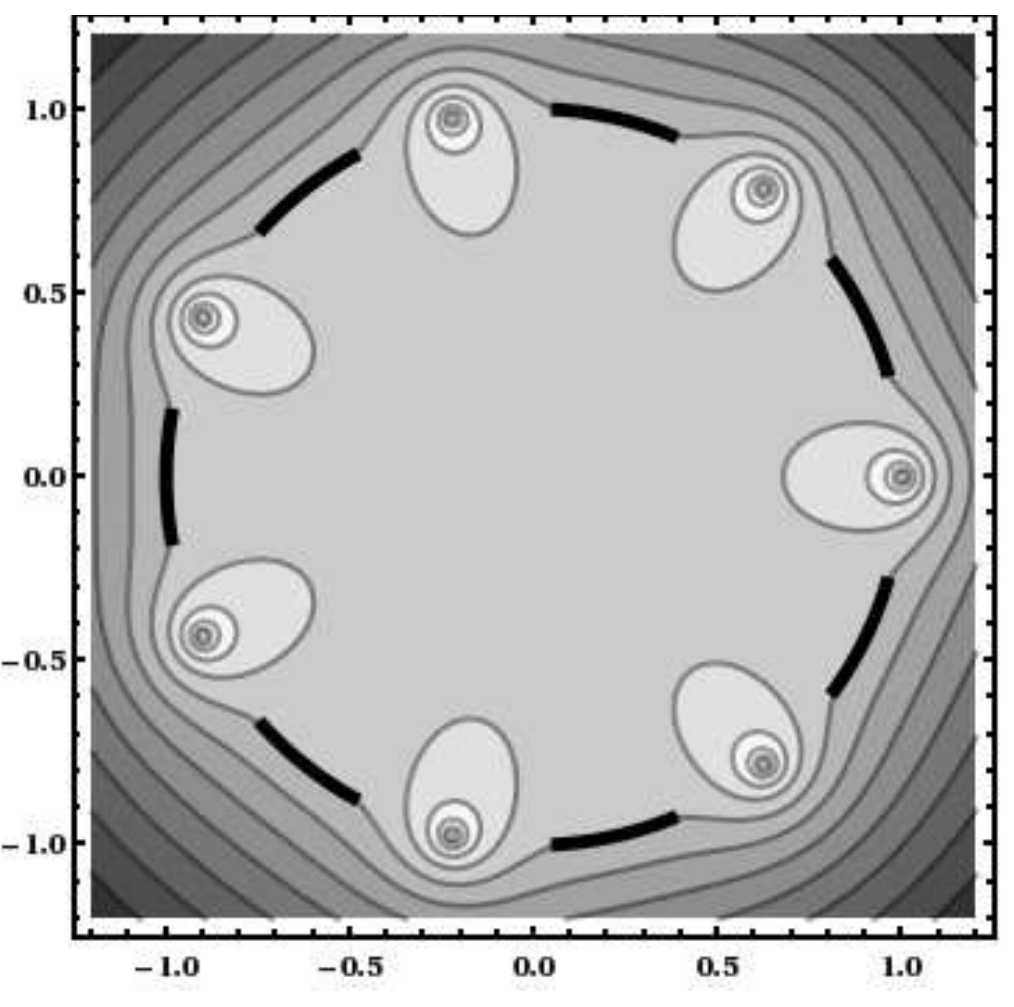}%
\includegraphics[width=0.5\textwidth]{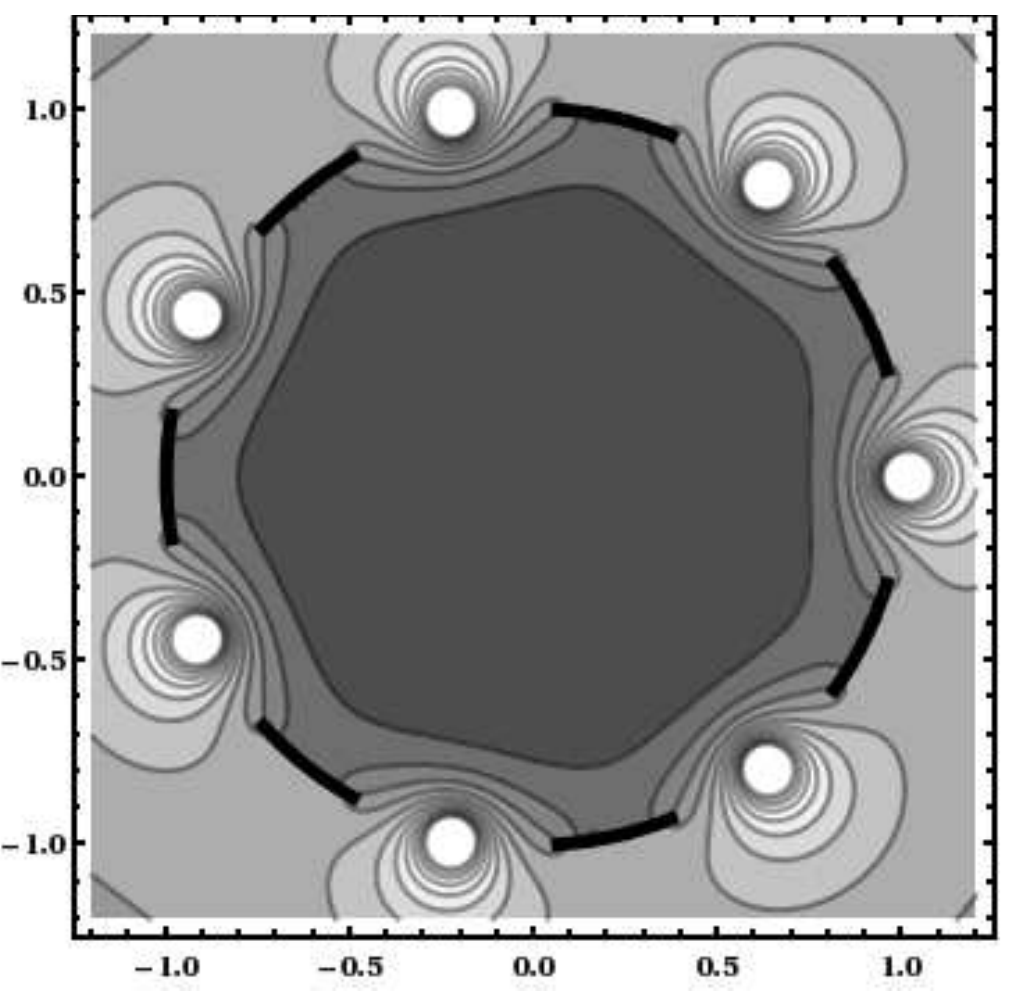}%
\caption{The potential $\re \, V^{(n{\rm pt})}(w)$ (left) and the absolute value of the electric field (right) for $n=7$ with $N/\xi=6$. The continuous charge distribution is denoted by the black arcs.}
\label{fig:npt}
\end{figure}

Let us consider a symmetric configuration (see Fig.~\ref{fig:npt}) where $n$ external particles have all equal charges
$\xi$ and are located at angles $\tau_a= 2\pi a/n$, where $a=1,\ldots,n$.
For a continuum charge equally distributed into $n$ symmetric blobs between the external charges, the potential can be immediately written in terms of the boundary one-point solution above. Namely,
\bea
 V^{(n{\rm pt}) }(w) &=& V^{\rm (1pt)}(w^n) \nn\\
 \rho^{(n{\rm pt}) }(t) &=& n\rho^{\rm (1pt)}(nt) \ .
\eea
We show the potential and the electric field for a choice of parameters in Fig.~\ref{fig:npt}.

The value $U$ of the potential at the conducting planes is unchanged, as well as the potential $\re \, V_c(w_a)$ felt by the external charges, which fixes $\mE_{c-\xi}$.
Notice that while conformal mappings conserve charge locally, in $w\mapsto w^n$ the total charge is multiplied by $n$ since there the mapping takes the charges to $n$ distinct images. Hence, we actually should start by a one-point solution with $N^{\rm (1pt)} = N/n \equiv \tilde N$. Consequently, the energy can be written as
\bea \label{eq:SE}
 \mE^{(n{\rm pt}) } &=& n \mE^{\rm (1pt)} + \mE_{\xi} \nn\\
                              &=& \frac{n}{2}\left[-2F(\xi)+ F(2\xi) - F(\tilde N) - F(\tilde N+2\xi)+ 2F(\tilde N+\xi)\right] + \mE_{\xi} \ ,
\eea
where the term arising from the mutual interactions of the external charges reads
\be
 \mE_{\xi} = -\frac{n\xi^2}{2}\sum_{a=1}^{n-1}\log\left|e^{2\pi i a/n}-1 \right| = -\frac{n\xi^2}{2} \log n \ .
\ee
Up to this additional term the result agrees with the approximation suggested in \cite{npt}, which was based on the limit $N \to \infty$ with $\xi$ fixed.

It is clear that the mapping $w \mapsto w^n$ can also be applied to more complicated solutions than the boundary one-point function. For example, we can map the antipodal configuration discussed above by, say, $w \mapsto w^2$, to obtain the asymptotics for a (quasi)-symmetric four-point partition function with two equal charges $\xi_1$ at $\tau=0$ and at $\tau=\pi$, and charges $\xi_2$ at $\tau=\pi/2$ and at $\tau=3\pi/2$. The total energy for this system becomes
\bea \label{eq:Etotansy}
 \mE &=&  2 \left.\mE^{\rm (2pt,a)}\right|_{N \to N/2} -\left(\xi_1^2+\xi_2^2\right)\log 2 \nn\\
 & = & 2 F(N/2+\xi_1+\xi_2) - 2 F(\xi_1+\xi_2) \nn\\
  &  & - \inv{2}\left[F(N/2+2 \xi_1+2\xi_2) +F(N/2+2 \xi_1) +F(N/2+2 \xi_2) +F(N/2)\right] \nn \\
       &   & + \inv{2}\left[ F(2 \xi_1+ 2 \xi_2)+F(2 \xi_1)+ F(2 \xi_2)\right] -\left(\xi_1+\xi_2\right)^2\log 2 \ ,
\eea
where the interactions of the external charges within themselves were treated separately.



\section{The bulk one-point function} \label{sec:b1pt}

We shall now point out a connection between the results of our previous article \cite{Jokela:2009fd} and the Poisson kernel.
Consider the bulk one-point partition function:
\bea \label{eq:Z2def}
 Z_{1+0}(w,\xi;N) &=&  \inv{N!}\int\left[\prod_{i=1}^{N}\frac{dt_i}{2\pi}|w-e^{it_i}|^{\beta\xi}\right]
 \left[\prod_{1\leq i<j\leq N}|e^{it_i}-e^{it_j}|^{\beta}\right] \nonumber \\
 &=& \left\langle \prod^N_{i=1} |1-we^{-it_i}|^{\beta \xi} \right\rangle_{{\rm CUE}_{\beta}(N)}
\eea
which equals the partition function of the Coulomb gas on the circle with an external charge $\xi$ at a fixed complex position $w$.\footnote{Notice that by rotational symmetry $Z_{1+0}$ is independent of $\arg w$.}
When $\xi=-(N-1)-2/\beta$, the integrand is a special case of the Poisson kernel,
\be
 P(S) = \frac{1}{C} \left( \frac{\det (1-\bar{S}^\dagger \bar S)}{|(\det (1-\bar{S}^\dagger S))|^2}
 \right)^{\beta (N-1)/2 +1} \ ,
\ee
when the matrix average is proportional to the unit matrix, $\bar{S} = w {\bf 1}_N$, and when the random
matrices $S$ belong to the CUE. In condensed matter physics, this case is used as a model for transport across
a disordered cavity, connected to the outside by non-ideal leads \cite{Brouwer} (see \cite{Beenakker} for
a general review). Here $N$ denotes the number of scattering channels. In this special case, the probability distribution function becomes
\be
 p(\{t_i\}) = \frac{1}{Z_{1+0} N!} \prod_{i=1}^{N}\frac{dt_i}{2\pi}|1-we^{-it_i}|^{-\beta (N-1) -2}
 \prod_{1\leq i<j\leq N}|e^{it_i}-e^{it_j}|^{\beta} \ ,
\ee
where we recognize the normalization factor $Z_{1+0}$ as (\ref{eq:Z2def}). In this case, when $\beta=2$,
$\xi = -N$ is a negative integer, $0<w=r<1$, (\ref{eq:Z2def}) has been calculated\footnote{Note that \cite{CL} assumed
$N\ge |\xi|$ but this condition is automatically satisfied with the above choice of $\xi$.} in \cite{CL}, with the simple result
\be
  Z_{1+0}(r,\xi;N) = (1-r^2)^{-\xi^2} \ .
\ee

In the case of positive $\xi$,  (\ref{eq:Z2def}) is much more complicated to calculate. In the double scaling limit we have obtained the 
result \cite{Jokela:2009fd},
\be\label{res}
 Z_{1+0}(r,\xi;N) = e^{-2\mE + {\cal D}(N)} \ ,
\ee
where the exponent is
\bea \label{exponent}
 \mE &=& \theta(r_c-r) \frac{\xi^2}{2} \log\left(1-r^2\right) \\
& &+ \theta(r-r_c) \left[  - \frac{(N+2 \xi)^2}{4}\log
\frac{1+\chi}{1+\delta(r)}-\frac{N^2}{4}\log
\frac{1-\chi}{1-\delta(r)} \right. \nonumber \\
& & \left. \ \ \ \ \ \ \ \ \ \ \ \ \ +\frac{\xi^2}{2}\log\frac{4\chi}{(1+\delta(r))^2}
 \right] \nonumber \ ,
\eea
where
\be\label{eq:rc}
r_c=\frac{N}{N+2\xi} \ ; \
\delta(r) = \frac{1-r}{1+r} \ ; \ \chi= \frac{\xi}{N+\xi} \ ,
\ee
and the term ${\cal D}(N)$ includes contributions from subleading terms, in particular from
terms of $\morder{N^0}$.
The function in the exponent has two possible forms, depending on the location of the bulk external charge. If it is sufficiently close
to the unit circle ($r\rightarrow 1$), a gap  $[-t_c,t_c]$ opens up in the continuous charge density $\rho(t)$, with the size $2t_c$ given by
\be
 \cos^2\frac{t_c}{2} = \frac{1-\chi^2}{1-\delta(r)^2} \ ,
\ee
where $\chi= \xi/(N+\xi)$. Thus $t_c$ vanishes at $r=r_c$ which signals the transition between gapped and ungapped charge distributions, reflected
in the form of (\ref{exponent}).

The result (\ref{res}) for positive $\xi$ can be applied to calculate the cumulative distribution function for long lattice paths in the discrete
polynuclear growth model, as we will discuss next.

\section{Bulk $n$-point functions and polynuclear growth}\label{sec:bulknpt}

\begin{figure}[h!bt]
\begin{center}
\noindent
\includegraphics[width=0.7\textwidth]{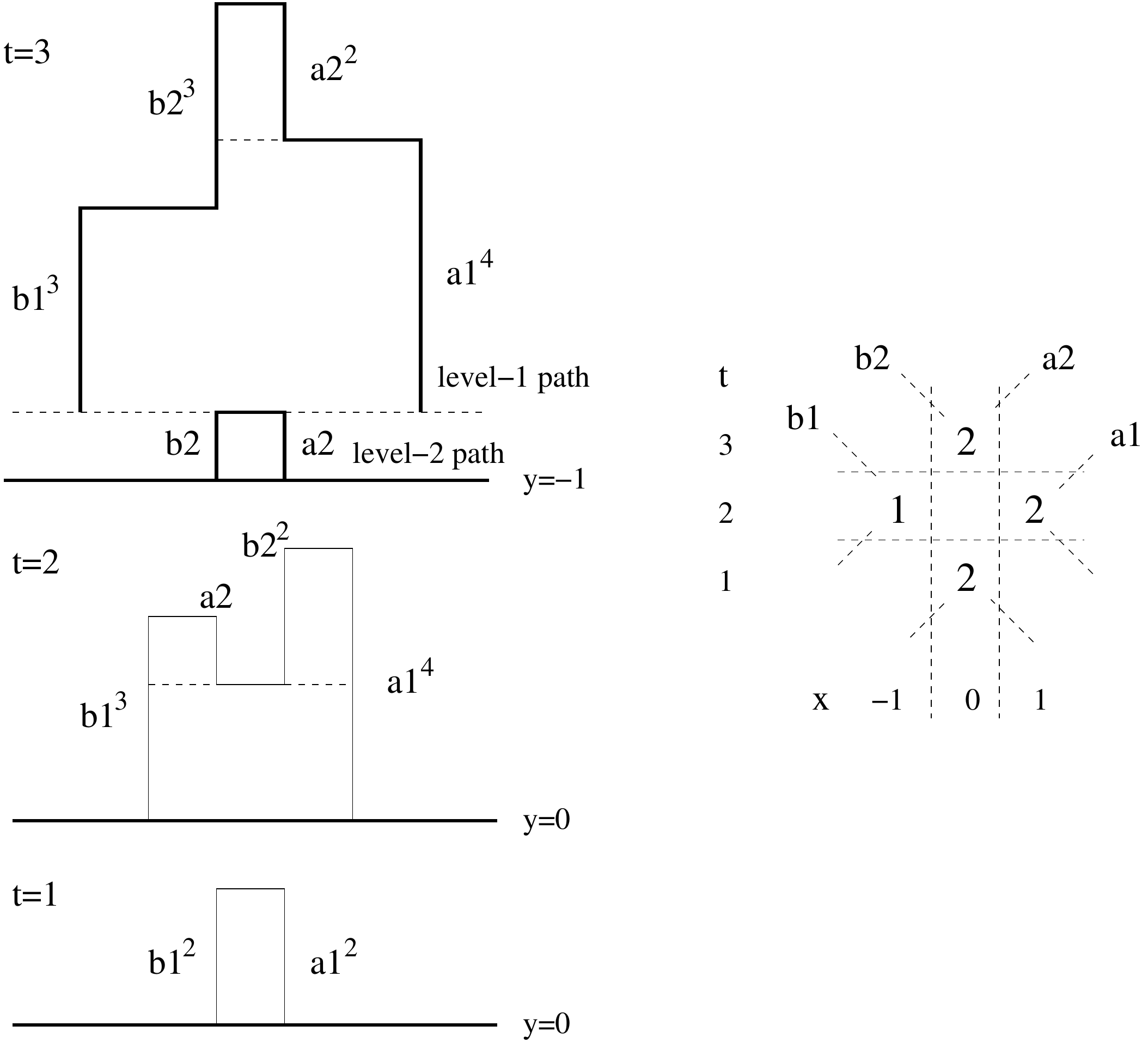}
\end{center}
\caption{A sketch of mapping a non-negative integer matrix to a box diagram according to the
rules of the discrete polynuclear growth (compare with \cite{Forrester}).}
\label{fig5}
\end{figure}

Next we study the gas with $n$ external charges in the bulk.
Recall that the partition function (\ref{eq:Idef}) becomes the CUE expectation value (we repeat the equation (\ref{eq:CUEexp}) for convenience),
\be\label{eq:apu}
Z_{n}(\beta;\{\xi_a\};N)  = \prod_{a<b} |w_a-w_b|^{\beta \xi_a\xi_b}\cdot
 \left\langle \prod^n_{a=1} \prod^N_{i=1} |1-w_ae^{-it_i}|^{\beta \xi_a} \right\rangle_{{\rm CUE}_\beta (N)} \ .
\ee
We point out that \eq{apu} is related to the weight of a lattice path in the discrete polynuclear growth model.
We will follow the presentation of the model in \cite{png} or the review \cite{Forrester} where more
details can be found. Here we just summarize
the basic idea. One starts from an ensemble of $m\times m$ weighted non-negative integer matrices $X=(x_{ij})$. Each entry of the matrix carries
a weight factor $(1-a_ib_j)(a_ib_j)^{x_{ij}}$, interpreted as coming from a geometric distribution (hence the normalization factor $(1-a_ib_j)$).
One then rotates the matrix counterclockwise by $45^{\circ}$, and starts reading the entries from the bottom corner, and growing a box diagram
following a set of rules. Basically one starts from the $y=0$ ground level and keeps adding rows of vertical boxes, the height of each box given by an entry of the matrix. The
addition of a box is a ``nucleation event''. The
matrix entries are read one horizontal row at a time starting from the bottom corner. At each step, the previously
constructed part of the diagram grows wider by one step to both left and right, before the new boxes are added on the top. Thus the box diagram will grow into a pyramid shape.
The left and right sides of the boxes carry weights
$b^{x_{ij}}$ and $a^{x_{ij}}$, respectively. As the boxes grow wider, they will sometimes overlap, in which case
 the overlapping part of the diagram is moved to a new sub-diagram below the ground level, $y<0$. In the end, when the box diagram is complete, the left and right edges of the profiles of the main diagram and subdiagrams
 are interpreted as pairs of weighted
non-intersecting lattice paths of $2m-1$ steps. The profiles with endpoints at $y=-l+1$, $l=1,\ldots ,m$ are called
level-$l$ paths. See Fig.~\ref{fig5} for a sketch of the process (following \cite{Forrester}).

\begin{figure}[ht]
\begin{center}
\includegraphics[width=0.5\textwidth]{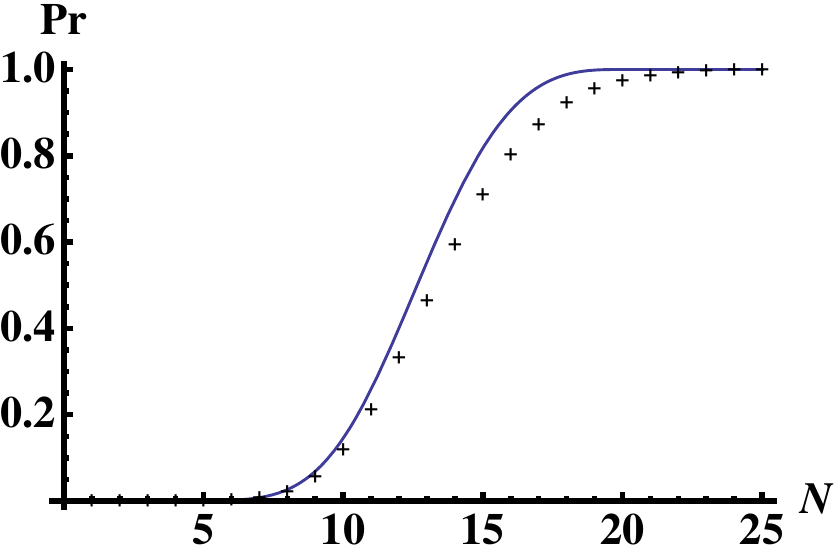}%
\includegraphics[width=0.5\textwidth]{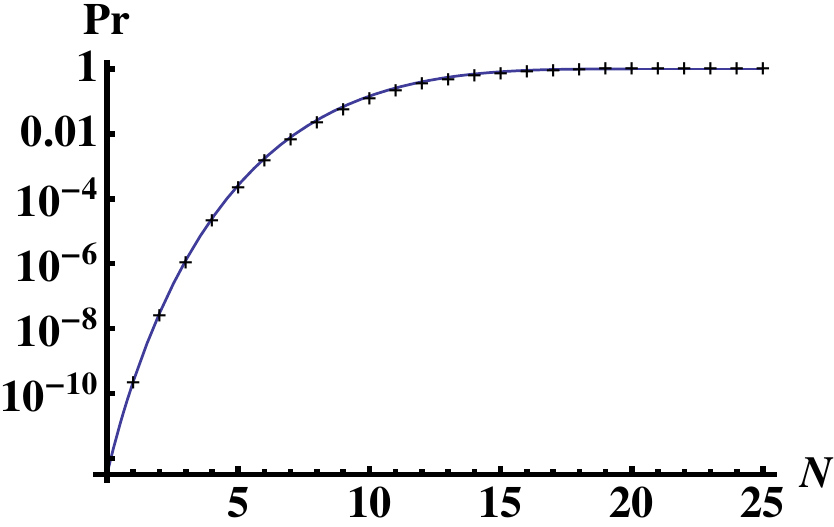}
\end{center}
\caption{The cumulative probability distribution function (\ref{eq:PNGres}) in the scaling limit (solid line) compared to the exact result (crosses).}
\label{fig:Pr}
\end{figure}

An interesting random variable in the polynuclear growth model is the maximum height of the level-1 path
(from the profile of the main part of the diagram, starting from and ending at the ground level $y=0$),
denoted $h^{\square}$. A fundamental
statistical quantity is its cumulative probability density ${\rm Pr}(h^{\square}\leq N)$. This is where the connection to circular unitary
ensemble can be established. One can show that
\be\label{eq:PNG}
 {\rm Pr}(h^{\square}\leq N) = \prod^m_{i,j} (1-a_ib_j)\cdot
 \left\langle \prod^m_{j=1} \prod^N_{k=1} (1-a_je^{it_k})(1-b_je^{-it_k}) \right\rangle_{{\rm CUE}_2 (N)} \ .
\ee
Now consider the special case of left-right symmetric weighting, $b_j=a_j$. Furthermore, suppose that $m=m_1+m_2+\cdots +m_n$ for some $n$, and make the following special choice for the weight factors,
\be
a_1=\cdots = a_{m_1} \equiv r_1 ,\  a_{m_1+1}=\cdots = a_{m_1+m_2} \equiv r_2,\ldots ,\ a_{m-m_n+1}=\cdots = a_{m} \equiv r_n  \ .
\ee
Then (\ref{eq:PNG}) becomes
\be\label{eq:PNG2}
 {\rm Pr}(h^{\square}\leq N) = \prod^n_{a,b} (1-r_ar_b)^{m_am_b}\cdot
 \left\langle \prod^n_{a=1} \prod^N_{k=1} |1-r_ae^{-it_k}|^{2m_a} \right\rangle_{{\rm CUE}_2 (N)} \ .
\ee
We recognize this as a special case of (\ref{eq:apu}) with
\be
 \beta = 2\ ; \  \xi_a = m_a \ ; w_a = r_a \ , \  a=1,\ldots ,n \ ,
\ee
adjusting the normalization factors. Thus, the bulk $n$-point functions compute the cumulative probability densities of the level-1 paths in the discrete PNG model. Note that a bulk two-point function with $r_1=0, r_2=r$ formally becomes equal to the bulk one-point function, which we discussed in the previous section.
In this case PNG becomes independent of $\xi_1=m_1$, and denoting $\xi_2=m_2\equiv m$ we find
\be\label{eq:PNG3}
 {\rm Pr}(h^{\square}\leq N) = (1-r^2)^{m^2}\cdot
 \left\langle \prod^N_{k=1} |1-re^{-it_k}|^{2m} \right\rangle_{{\rm CUE}_2 (N)} = (1-r^2)^{m^2} Z_{1+0}(r,m;N) \ .
\ee
The expectation value is the one-point partition function (\ref{eq:Z2def}) with $w=r,\beta=2,\xi=m$. In the double scaling
limit $N,m\gg 1$ with $N/m$ fixed, we derived and approximate result (\ref{res}), so the
cumulative distribution function becomes
\be\label{eq:PNG4}
 {\rm Pr}(h^{\square}\leq N) = (1-r^2)^{m^2} e^{-2\mE } \ ,
\ee
with the exponent (\ref{exponent}) and up to the subleading corrections ${\cal D}(N)$ in (\ref{res}) which we shall neglect. We expect
that form to be valid even if we keep $m$ fixed and vary $N$, as long as $N,m\gg 1$. In the growth model,
 the weight factor $r$ has a fixed chosen value. On the other hand, for the cumulative distribution
  function, we need to establish that ${\rm Pr}(h^{\square}\leq N) \rightarrow 1$ as $N \rightarrow \infty$.
Since $r$ is fixed, at some point it will become less than the critical value $r_c=N/(N+2m)$, when $N$ increases.
When $r<r_c$, the exponent $\mE$ becomes
\be
\mE = \frac{m^2}{2} \log\left(1-r^2\right) \ ,
\ee
which is independent of $N$. We interpret this as a sign that the cumulative probability distribution has
saturated, indeed (\ref{eq:PNG4}) becomes equal to one (up to higher order corrections).
Thus ${\rm Pr}(h^{\square}\leq N) \rightarrow 1$
as $N\rightarrow N_c$, where
\be
N_c=\frac{2mr}{1-r} \ ,
\ee
found from (see (\ref{eq:rc}))
\be
  r = \frac{N_c}{N_c +2m} \ .
\ee
This means that $N_c$ is a probabilistic upper bound for the height of level-1 paths, in the sense that
cumulative distribution function saturates. For $N<N_c$, the cumulative distribution function is
\bea\label{eq:PNGres}
{\rm Pr}(h^{\square}\leq N)  &=& (1-r^2)^{m^2}
 \left(\frac{1+\chi}{1+\delta (r)}\right)^{(N+2m)^2/2} \nonumber \\
&& \times \left(\frac{1-\chi}{1-\delta (r)}\right)^{N^2/2}
\left(\frac{4\chi}{(1+\delta (r))^2}\right)^{-m^2} \ .
\eea
Fig.~\ref{fig:Pr} shows an example plot for $m=10,r=0.5$ (so that $N_c=20$) in linear and logarithmic scales. We also show the exact probabilities which are easy to obtain from Eq.~\eq{PNG3} for not too large $N$ by evaluating numerically the corresponding Toeplitz determinant
\be
 Z_{1+0}(r,m;N) = \det T_N(h)
\ee
for the function $h(t) = |1-re^{-it}|^{2m}$ [compare to Eq.~\eq{Zexc}].

\subsection{Lattice path interpretation}

The bulk $n$-point functions at $\beta=2$ and with charges on the real line also admit a lattice path interpretation which is very similar to that of the boundary one-point function discussed in Section~\ref{sec:lpi}.
Actually, we may again write a connection to  $N$ non-intersecting single move $ld/rd$ (left diagonal/right diagonal) paths with $2m = 2\sum_a m_a$ steps with the same starting and ending locations as for the boundary function (see Fig.~\ref{fig:vicwalk}):
\bea \label{eq:npint}
 & &\left[\prod_{a<b}|r_a-r_b|^{-2 m_a m_b}\right] \cdot Z_{n+0}(\{r_a\},\{m_a\};N) = \left\langle \prod^N_{k=1} \prod_{a=1}^n |1-r_ae^{-it_k}|^{2m_a} \right\rangle_{{\rm CUE}_2 (N)} \nn\\
 && = G^{ld/rd}_{2m} (\{ l^{(0)}_j=2(N-j)\}_{j=1,\ldots,N};\{ l_j=2(N-j)\}_{j=1,\ldots,N}) \ ,
\eea
but now we need to add some nontrivial weights to the paths (see \cite{Forrester}, proposition 8.13). We denote the weights of the left (right) diagonal paths at step $k$ by $w_k^\mp$. The weights for the $m$ first steps are
\bea
 w_1^- &=& \cdots \ = w_{m_1}^- = r_1 \nn\\
 w_{m_1+1}^- &=& \cdots \ = w_{m_1+m_2}^- = r_2 \nn\\
 &\vdots& \nn\\
 w_{m-m_n+1}^- &=& \cdots \ = w_m^- =r_n \nn\\
 w_1^+ &=& \cdots \ = w_m^+ = 1
\eea
while for the remaining $m$ steps we flip $+ \leftrightarrow -$, \ie,
\be
 w_{k+m}^\mp = w_k^\pm \ ; \qquad k=1, \ldots , m \ .
\ee
According to \eq{npint} the $n$-point partition function is then proportional to the total weight of all $N$ non-intersecting $ld/rd$ paths with above boundary conditions and weights. Letting all $r_n \to 1$ we recover the result for the boundary one-point function (with charge $m$) as all the weights become equal to one. Notice also that for the $n=1$ case we can immediately extract the leading behavior of the total weight $G^{ld/rd}_{2m}$ in the double scaling limit by using Eqs.~(\ref{res}) and~(\ref{exponent}).
It would be interesting to know what other interpretations of the bulk $n$-point function (\ref{eq:Z2def}) exist for other values of $\beta \neq 2$ and noninteger $\xi_a$.


\bigskip
\noindent

{\bf \large Acknowledgments}

We thank P. Forrester for bringing the reference \cite{BNR} to our
attention.
N.J. has been supported in part by the Israel Science Foundation under grant no. 392/09 and in part at the Technion
by a fellowship from the Lady Davis Foundation. M.J. has been supported in part by the Villum Kann Rasmussen Foundation. E.K-V. has been supported in part by the Academy of Finland grant number
1127482. This work has also been supported in part by the EU 6th Framework Marie Curie Research and Training network ``UniverseNet''
(MRTN-CT-2006-035863).

\appendix

\section{Partition function to next-to-leading order} \label{app:ooN}

Consider the two-dimensional Coulomb gas on the circle, subject to an external potential $\xi V_{\rm ext}(z)$. The system is defined by the Hamiltonian
\be
 H = - \sum_{i,j=1}^N \log\left|e^{it_i}-e^{it_j}\right| + \xi \sum_{i=1}^N V_{\rm ext}\left(e^{it_i}\right) \ ,
\ee
where the positions of the unit charges are $z_i=e^{it_i}$. The partition function is then
\be
 Z(\beta,\xi,N) = \inv{N!}\int \prod_i \frac{dt_i}{2\pi} e^{-\beta H} \ .
\ee
We consider the limit where $N,\xi \to \infty$ with $N/\xi$ and $V_{\rm ext}$ fixed. This is a generalization of the limit which was discussed in the main text to an arbitrary external potential and general $\beta$. The aim of this Appendix is to sketch how to calculate the first two coefficients in the series (\ref{eq:logZexp})
where the coefficients depend on $\xi/N$ and may also include logarithms of $N$. In particular, we show how to relate $C_0$ to the electrostatic energy in the continuum limit. We shall do a direct brute-force calculation of the coefficient $C_1$, which is shown to have a simple structure. A more sophisticated method could use an effective theory of continuous charge distributions \cite{Wiegmann:2005eh}.

Recalling the saddle-point approach (\ref{saddlept}), we write first
$ \log Z(\beta,\xi,N) = -\beta H_0 + \Delta_f$, 
where $H_0$ is the (global) minimum value of the Hamiltonian\footnote{This step may be subtle if the external potential is complicated, and $H$ has several separate minimum configurations. This can occur if the support of the charge density is disconnected in the large $N$ limit.
One can check that our analysis remains valid when global minimum configurations are used both in the discrete and continuous systems. Actually, we will not use the results of this Appendix in cases where this subtlety appears.}
and $\Delta_f$ is the contribution from fluctuations around the minima.
For nonzero external potential, it is hard to calculate $H_0$ explicitly. Hence we introduce a further correction term by writing $-\beta H_0 = -\beta \mE + \Delta_d$, where $\mE$ is the electrostatic energy in the continuum limit, and $\Delta_d$ is a discretization correction. The continuum energy $-\beta \mE$ scales as $N^2$ and gives exactly the leading term $C_0 N^2$, while the next-to-leading coefficient $C_1 N$ can be extracted from the various correction terms $\Delta$. We shall calculate the discretization correction at \order{N} by further dividing it to two separate contributions: one arising from ``self-energies'' of the point charges, and one from differences in near-neighbor interaction strengths. Hence we have, at \order{N},
\be
 \log Z(\beta,\xi,N) = -\beta \mE + \Delta_f + \Delta_{\rm se} + \Delta_{\rm nn} \ .
\ee
For more precise definitions see the analysis below, where we discuss the various terms one by one.

Notice also the equilibrium charge densities of the continuous and discrete systems may be different at next-to-leading order in $1/N$.
One could naively expect such a deformation to cause an additional \order{N} correction term.
However, recall that $\mE$ is minimized by the leading order charge density of the continuous system. Therefore the next-to-leading order change of the total energy due to \emph{any} perturbation of the charge density vanishes.

In summary, using the results for the correction term $\Delta$ from Eqs. \eq{Dfres}, \eq{Dseres}, and \eq{Dnnres} below, we obtain the leading term $C_0 N^2 = -\beta \mE$ as in (\ref{lead}), followed
by 
\bea \label{eq:ooNcorr}
 C_1 N &=&  \Delta_f + \Delta_{\rm se} + \Delta_{\rm nn} + \morder{N^0} \nn \\
 &=& \frac{\beta\!-\!2}{2}\!\!\int dt \rho(t) \log 2\pi\rho(t) +\left[\frac{2\!-\!\beta}{\beta}+\frac{\beta}{2}\log\frac{\beta}{2}-\log\Gamma\!\left(\frac{\beta}{2}\!+\!1\right)\right] N \ ,
\eea
where the total energy of the continuum system in equilibrium $\mE$ and the corresponding charge density $\rho(t)$ can in principle be calculated by using the methods of Section~\ref{app:genmet}. Notice that $C_1$ vanishes for $\beta=2$.

\subsection{Dyson gas}

To motivate the calculation of $\Delta_f$ below,
we consider the case $V_{\rm ext} \equiv 0$, \ie, the Dyson gas \cite{Dyson:1962es}. Even though we will be interested in the limit of large $N$ with $\beta$ fixed, it is illustrative to consider the low temperature behavior ($\beta \to \infty$) first. The partition function is given by
\bea
 Z(\beta,N) &=& \inv{N!} \int  \prod_i \frac{dt_i}{2\pi} \prod_{i<j} \left|e^{it_i}-e^{it_j}\right|^\beta \ .
\eea
In this case $C_0=0$ (consistently with the expectation from electrostatics).
For large $\beta$, the partition function may be calculated by using a saddle-point approach. By rotational symmetry, we fix $t_1=0$. There are $(N-1)!$ equilibrium configurations, given by the permutations of the configuration $t_j = 2 \pi (j-1)/N$ with $t_1$ fixed. The value of the integrand at the equilibrium is given by
\be
 \sum_{j<k} \left|e^{2\pi i (j-1)/N}-e^{2\pi i (k-1)/N}\right|^\beta = N^{\frac{\beta N}{2}} \ .
\ee
Because of symmetry we may fix $0=t_1<t_2<\cdots<t_N<2\pi$ and expand around one of the minima. By using
\be
 M_{jk} \equiv \frac{\partial^2}{\partial t_j \partial t_k} H = \left\{\begin{array}{rl} -\sum_{\ell \ne j} \frac{1}{\left|e^{it_j}-e^{it_\ell}\right|^2} \ ; & \qquad j=k \\
 \frac{1}{\left|e^{it_j}-e^{it_k}\right|^2} \ ; & \qquad j \ne k \end{array}\right.
\ee
we can write
\bea  \label{eq:Zfluct}
 Z(\beta,N) &=& \frac{1}{N}  N^{\frac{\beta N}{2}} \int_V \prod_{j=2}^N\frac{ds_j}{2\pi} e^{-\beta \Delta H\left(\{s_j\}\right)} \\ \label{eq:Zsaddl}
 &=& \frac{1}{N}  N^{\frac{\beta N}{2}} \int_V \prod_{j=2}^N\frac{ds_j}{2\pi} e^{-\frac{\beta}{2} \sum_{j,k} s_j M_{jk} s_k} \left[1+\morder{\beta N^3 s_j^3}\right] \ ,
\eea
where $s_j = t_j - 2\pi(j-1)/N$ are the deviations of $t_j$ from the equilibrium configuration, $\Delta H$ is the Hamiltonian minus its value at equilibrium, the matrix $M$ is understood to be evaluated at the equilibrium point, and the region of integration $V$ is fixed by $0<t_2<\cdots<t_N<2\pi$. 

In the end we will consider the behavior of $Z$ at large $N$. Therefore, we will analyze the dependence of the various terms in Eq.~\eq{Zsaddl} on $N$ in addition to their $\beta$ dependence. 
For large $N$ the dominant contribution to the matrix $M$ comes from the charges within distances $\left|e^{it_j}-e^{it_k}\right| \sim 1/N$. Therefore, the (near diagonal) elements of the matrix $M_{jk}$ behave as $M_{jk} \sim N^2$. The integral \eq{Zfluct} is dominated by the region where the argument of the exponential function is \order{1}. Thus, for large $\beta$ (and $N$), the variables scale as $s_j \sim 1/(\sqrt{\beta} N)$. The boundaries of $V$ lie at $s_j \sim 1/N$. Therefore, for large $\beta$ the saddle-point approximation works, and we find
\be
 Z(\beta,N) = N^{\frac{\beta N}{2}} \inv{N (2\pi\beta)^{(N-1)/2}\sqrt{\det M}} \left[1+\morder{\inv{\beta}}\right] \ .
\ee
Notice that the \order{\inv{\sqrt{\beta}}} correction term  vanishes in the $s_k$ integration since the integral is odd. The \order{\inv{\beta}} term might have a sizeable (polynomially) $N$ dependent coefficient. 

If we take $N \to \infty$ the determinant $\det M$ may also be evaluated.
In this limit $M$ becomes a Toeplitz matrix:
\be
 \frac{4\pi^2}{N^2}M_{jk} \simeq \left\{\begin{array}{rl} -2 \zeta(2)\ ; & \qquad j=k \\
 \frac{1}{(j-k)^2} \ ; & \qquad j \ne k  \quad \mathrm{with}\quad  (j-k)^2 \ll N^2 \ . \end{array}\right.
\ee
More precisely, the elements near the diagonal, which give the dominant contribution to the asymptotics, approach the Toeplitz form.\footnote{Notice that
since the charges lie on a circle, the distance $|j-k|$ needs to be understood modulo the matrix dimension, \ie, the elements near the upper right and lower left corners of the matrix are also relevant. This is also required for $M$ to be asymptotically of the Toeplitz form.}
The asymptotics of $\det M$ may be solved by using Szeg{\"o}'s limit
theorem for Toeplitz determinants. We get
\be
 \det M = \left( \frac{N}{2\pi}\right)^{2N} e^{N\left(\log 2\pi^2 -2\right) + \morder{\log N}} \ ,
\ee
and consequently, taking first $\beta \to \infty$ and then $N \to \infty$,
\be \label{eq:logZsaddl}
 \log Z(\beta,N) = \frac{\beta -2}{2} N\log N + N\left(\frac{1}{2}\log \beta -\inv{2}\log \pi + 1\right) +
 \morder{\beta^0\log N} + \morder{\frac{1}{\beta}} \ ,
\ee
where the \order{\inv{\beta}} term depends polynomially on $N$.
Now, let us compare this to the exact result
\bea
 \log Z(\beta,N) &=& \log\frac{\Gamma\left(\frac{\beta N}{2}+1\right)}{N! \Gamma\left(\frac{\beta}{2}+1\right)^N} \nonumber \\
  &=& \frac{\beta -2}{2} N\log N +\left[\frac{\beta\!-\!2}{\beta}\left(\log 2\pi\!-\!1\right)+\frac{\beta}{2}\log\frac{\beta}{2}-\log\Gamma\!\left(\frac{\beta}{2}\!+\!1\right)\right] N \nonumber \\
  &&+ \inv{2}\log\frac{\beta}{2} + \morder{\inv{N}} \label{eq:Zasex}
\eea
For large $N$ and $\beta$ one may check that \eq{logZsaddl} is indeed correct. The correction terms are found to be $\morder{N^0}+ \morder{\frac{N}{\beta}}$ so that  
the \order{\log N} terms actually cancel.

Let us then discuss the limit $N \to \infty$ with fixed $\beta$. In this limit the saddle-point approximation breaks down. However, we
may extract the result for the integral in Eq.~\eq{Zfluct} in this limit by comparing\footnote{Interestingly, by doing the comparison instead in the limit $\beta \to \infty$ with $N$ fixed we find the \emph{exact} result $\det M = 2 \Gamma(N)^2/\left(N 2^N\right)$.} to \eq{Zasex},
\bea \label{eq:Ifdef}
 I_f&\equiv&\int_V \prod_{j=2}^N\frac{ds_j}{2\pi} e^{-\beta\Delta H\left(\{s_j\}\right)} \nonumber\\
 &=& \exp\left\{-N\log N +\left[\frac{2-\beta}{\beta}+\frac{\beta}{2}\log\frac{\beta}{2}-\log\Gamma\left(\frac{\beta}{2}+1\right)\right] N + \morder{N^0}\right\} \ .
\eea
This result will be useful in the calculation for general $V_{\rm ext}$ below.

Let us conclude the saddle-point analysis of the partition function by an important observation.
From above calculation we learn that only variables $s_j$, $s_k$ with $|j-k|\ll N$ were coupled in the saddle-point integration, as seen from the structure of $M$. That is, only the near-neighbor interactions are relevant when fluctuations around the extremal configurations are considered at large $N$. While we were not able to calculate the integral for general $\beta$, this statement remains true independently of $\beta$, and also in the presence of an external potential, as we shall sketch below.

\subsection{Fluctuation corrections}

The treatment of the term $\Delta_f$ is the most tricky one, and we will only sketch how to obtain the result. In analogue to \eq{Zfluct} we may write
\be
  Z(\beta,\xi,N) = e^{-\beta H_0} \int_V \prod_{j=1}^N\frac{ds_j}{2\pi} e^{-\beta \Delta H\left(\{s_j\}\right)} \ ,
\ee
where $\Delta H$ is again the difference between the Hamiltonian and its equilibrium value, and $V$ is defined by $t_1<t_2<\cdots<t_N<t_1+2\pi$. We assume that $V_{\rm ext}$ breaks the rotational symmetry, and hence fixing $t_1$ is neither possible nor necessary. For sufficiently large $N$, the equilibrium condition can be described by a continuous charge distribution. Let us denote the density by $\hat \rho$ and assume normalization $\int \frac{dt}{2 \pi} \hat\rho (t) = 1$. Then $\hat \rho \equiv 1$ in the absence of external potential. Let us do the change of variables
\be \label{eq:subst}
 u_j = \hat \rho(t_j) s_j \equiv \hat \rho_j s_j \ .
\ee
We obtain
\be \label{eq:Dfint}
 e^{\Delta_f} = \inv{\prod_j \hat\rho_j} \int_V \prod_{j=1}^N\frac{du_j}{2\pi} e^{-\beta \Delta H\left(\{u_j\}\right)} \ .
\ee
The claim is now, that the remaining integral becomes $I_f$ of \eq{Ifdef} for large $N$. We do not have a precise proof (except for large $\beta$ where the saddle-point method works), but one can understand why the result arises. As noted above for $V_{\rm ext}=0$, only near-neighbor interactions matter when fluctuations around the minima of the Hamiltonian are considered. Locally, within distances \order{1/N}, in the minimum configuration the distances of adjacent charges are to first approximation constant, and equal to $2\pi/(N \hat\rho)$. Hence the substitution \eq{subst} removes dependence of the Hamiltonian on $\hat \rho$ for interactions within distances \order{1/N}. At \order{N} in $\Delta_f$ we may replace $\Delta H$ in \eq{Dfint} by its Dyson gas counterpart in $I_f$ of Eq. \eq{Ifdef}, and obtain
\bea \label{eq:Dfres}
 \Delta_f &\simeq& -\sum_j \log \hat\rho_j + \log I_f \nn\\
 &=&-\int dt \rho (t) \log 2\pi\rho(t) +\left[\frac{2-\beta}{\beta}+\frac{\beta}{2}\log\frac{\beta}{2}-\log\Gamma\left(\frac{\beta}{2}+1\right)\right] N \ ,
\eea
where the term involving the charge density could be written as an integral over the continuum charge density up to higher order corrections.
Notice that the dependence on the external potential appears only through the modification of the charge density.

The above arguments which lead to the result \eq{Dfres} may be illustrated by considering the first term in the saddle-point calculation, which is the leading one for large $\beta$. The matrix $M_{jk}$ is found by taking the second derivative of the Hamiltonian and evaluating it at the equilibrium. The external potential appears explicitly only in the diagonal elements, which read
\be
 M_{jj} = -\sum_{\ell \ne j} \frac{1}{\left|e^{it_j}-e^{it_\ell}\right|^2} + \xi \frac{\partial^2}{\partial t_j^2} V_{\rm ext}(t_j) \ .
\ee
The $\xi$-dependent term is suppressed by $1/N$ with respect to the near-neighbor contributions, as expected, and can be neglected. Therefore the approximate result for $M_{jk}$, arising from near-neighbor terms, can be written in terms of $\hat \rho$ as
\be
 \frac{4\pi^2}{N^2}M_{jk} \simeq \left\{\begin{array}{rl} -2 \zeta(2) \hat \rho_j^2 + \morder{\inv{N}} \ ; & \qquad j=k \\
 \frac{\hat \rho_j \hat \rho_k }{(j-k)^2}  + \morder{\inv{N}} \ ; & \qquad j \ne k  \quad \mathrm{with}\quad  (j-k)^2 \ll N^2 \ . \end{array}\right.
\ee
At leading order we were able to use the geometric mean density $\sqrt{\hat \rho_j \hat \rho_k}$ for the near-diagonal terms.
By making the substitution \eq{subst} in the saddle-point integral, we confirm that dependence on $\hat \rho$ appears only through the change in the integration measure. Hence the saddle-point result is in agreement with \eq{Dfres}.

\subsection{Self-energy corrections}

The terms $\Delta_{\rm se}$ and $\Delta_{\rm nn}$ arise from the difference of the minimum energies of the continuum and discrete systems. We calculate them by taking the continuum system, by compressing suitable clumps of the continuous charge into point charges, and by studying the change in energy. The easiest term to calculate is  $\Delta_{\rm se}$, which arises from the self energies of the clumps that are absent in the discrete system. The self-energy of a single clump is given to first approximation by
\be
 E_{\rm se} = \frac{1}{2\delta^2} \int_{-\delta/2}^{\delta/2} dx dy \log|x-y| = -\frac{3}{4} +\inv{2}\log\delta\ ,
\ee
where $\delta = 2\pi/(N \hat \rho)$ is the size of the clump. Summing over the clumps gives the \order{N} result
\be \label{eq:Dseres}
 \Delta_{\rm se} = -\beta \sum_{j=1}^N E_{{\rm sc},j} = \frac{\beta N}{2}\log N - \frac{\beta N}{2}\log 2\pi + \frac{3 \beta }{4} N + \frac{\beta}{2}\sum_{j=1}^N \log\hat\rho_j \ .
\ee

\subsection{Near-neighbor interactions}

The term $\Delta_{\rm nn}$ may be calculated similarly as the self-energy corrections. The change in energy between two clumps is
\be
 \Delta E_{\rm nn} = \log d - \frac{1}{\delta^2} \int_{-\delta/2}^{\delta/2} dx dy \log|d+x-y|
\ee
where $d\ge\delta$ is the distance between the clumps. For $d\sim \delta \sim 1/N$, we find
\be
 \Delta E_{\rm nn} = \frac{3}{2}-\inv{2}(m-1)^2\log\left(1-\inv{m^2}\right)+2m\log\left(1+\inv{m}\right)
\ee
with $m=d/\delta$. Summing over all pairs of clumps gives
\be \label{eq:Dnnres}
 \Delta_{\rm nn} = \beta N \sum_{m=1}^\infty  \Delta E_{\rm nn} =  \frac{\beta N}{2}\log 2\pi - \frac{3 \beta }{4} N \ .
\ee
It is not hard to see that contributions for $d\gg 1/N$ are suppressed. This is signaled by the convergence of the sum over $m$ in the above calculation: the region $m \gtrsim N$ where our approximation for $\Delta E_{\rm nn}$ breaks down does not contribute at leading order.

\end{document}